\def\StructureTable{1}

\def\rn{}
\def\nn#1 #2{#2. #1}				% Name with 1 initial
\def\nnn#1 #2 #3{#2. #3. #1}			% Name with 2 initials
\def\nnnn#1 #2 #3 #4{#2. #3. #4. #1}		% Name with 3 initials
\def\nnnnn#1 #2 #3 #4 #5{#2. #3. #4. #5. #1}	% Name with 4 initials
\def\dualand{ and\hbox{ }}				
\def\multiand{, and\hbox{ }}				
\def\rf#1;#2;#3;#4;#5 {{\frenchspacing\par\rn#1, #3 {\bf #4}, #5 (#2). \par}}
\def\rg#1;#2;#3;#4;#5;#6 {{\frenchspacing\par\rn#1, #3 {\bf #4}, #5 (#2). \par}}
\def\rfproc#1;#2;#3;#4;#5;#6 {{\frenchspacing\par\rn#1 #2, in {\it #3}, ed. #4 (#5: #6)\par}}
\def\rfprocp#1;#2;#3;#4;#5;#6;#7 {{\frenchspacing\par\rn#1 #2, in {\it #3}, ed. #4 (#5: #6), p#7\par}}

\def\rg#1;#2;#3;#4;#5;#6 {\par\rn#1 #2, {\it #3}, {\bf #4}, #5 (``#6'') \par}
\def\rf#1;#2;#3;#4;#5 {\par\rn#1, {\it #3}, {\bf #4}, #5 (#2)\par}
\def\rfbook#1;#2;#3;#4;#5 {{\frenchspacing\par\rn#1, {\it #3} (#4: #5, #2)\par}}
\def\rfproc#1;#2;#3;#4;#5;#6 {{\frenchspacing\par\rn#1 #2, in {\it #3}, ed. #4 (#5: #6)\par}}
\def\rfprocp#1;#2;#3;#4;#5;#6;#7 {{\frenchspacing\par\rn#1 #2, in {\it #3}, ed. #4 (#5: #6), p#7\par}}
\def\rfprep#1;#2;#3 {{\par\frenchspacing\rn#1, #3 (#2)\par}}
\def\rfprepp#1;#2;#3 {{\par\rn#1 #2, #3\par}}

\def\zero{{\bf 0}}

\def\brack#1{\langle\hbox{#1}\rangle}

\def\etal{{\frenchspacing\it et al.}}
\def\ie{{\frenchspacing\it i.e.}}
\def\eg{{\frenchspacing\it e.g.}}
\def\etc{{\frenchspacing\it etc.}}

\def\cf{{\frenchspacing\it c.f.}}

\def\beq#1{\begin{equation}\label{#1}}
\def\eeq{\end{equation}}
\def\beqa#1{\begin{eqnarray}\label{#1}}
\def\eeqa{\end{eqnarray}}
\def\eq#1{equation~(\ref{#1})}
\def\Eq#1{Equation~(\ref{#1})}
\def\eqn#1{~(\ref{#1})}

\def\fig#1{Figure~\ref{#1}}
\def\Fig#1{Figure~\ref{#1}}

\def\Sec#1{Section~\ref{#1}}
\def\Sec#1{Section~\ref{#1}}

\def\spose#1{\hbox to 0pt{#1\hss}}
\def\simlt{\mathrel{\spose{\lower 3pt\hbox{$\mathchar"218$}}
     \raise 2.0pt\hbox{$\mathchar"13C$}}}
\def\simgt{\mathrel{\spose{\lower 3pt\hbox{$\mathchar"218$}}
     \raise 2.0pt\hbox{$\mathchar"13E$}}}
\def\simpropto{\mathrel{\spose{\lower 3pt\hbox{$\mathchar"218$}}
     \raise 2.0pt\hbox{$\propto$}}}

\def\ed{\end{document}}

\def\imp{\Rightarrow}

\def\och{\&}

\def\beq#1{\begin{equation}\label{#1}}
\def\eeq{\end{equation}}
\def\beqa#1{\begin{eqnarray}\label{#1}}
\def\eeqa{\end{eqnarray}}
\def\eq#1{equation~(\ref{#1})}
\def\Eq#1{Equation~(\ref{#1})}
\def\eqn#1{~(\ref{#1})}

\def\Reals{\mathbb{R}}
\def\Integers{\mathbb{Z}}

\def\a{{\bf a}}

\def\rvec{{\bf r}}

\def\R{{\bf R}}

\def\R{{\bf R}}

\def\vpsi{{\boldsymbol\psi}}

\def\union{\cup}

\def\Aut{\hbox{Aut}\,}

\documentclass[twocolumn,amsmath,nofootinbib]{revtex4} % For astro-ph
\usepackage{amsfonts,amsbsy,epsf,hypernat} % amsfonts defines mathbb used for \Reals, hypernat needed to prevent arXiv from expanding [1-5] to [1,2,3,4,5]
\begin{document}

\def\mit{1}
\def\penn{2}

\def\affilmrk#1{$^{#1}$}
\def\affilmk#1#2{$^{#1}$#2;}

\title{The Mathematical Universe\footnote{Partly based on a talk given at the symposium ``Multiverse and String Theory: Toward Ultimate Explanations in Cosmology''
held on 19-21 March 2005 at Stanford University and on the essay \cite{TownesBook}.}}

\author{
Max Tegmark
}
\address{Dept.~of Physics, Massachusetts Institute of Technology, Cambridge, MA 02139}

\date{Submitted to {\it Found. Phys.}  April 7 2007, revised September 6, accepted September 30}

\begin{abstract}
I explore physics implications of the {\it External Reality Hypothesis} (ERH) that there exists an external physical reality 
completely independent of us humans. I argue that with a sufficiently broad definition of mathematics, 
it implies the {\it Mathematical Universe Hypothesis} (MUH) that 
our physical world is 
an abstract mathematical structure.
I discuss various implications of the ERH and MUH, ranging from standard
physics topics like symmetries, irreducible representations, units, free parameters, randomness and initial conditions
to broader issues like consciousness, parallel universes and G\"odel incompleteness.
I hypothesize that only computable and decidable (in  G\"odel's sense) 
structures exist, which alleviates the cosmological measure problem and may help 
explain why our physical laws appear so simple.
I also comment on the intimate relation between mathematical structures, computations, simulations and physical systems.
\end{abstract}


  
\maketitle

\setcounter{footnote}{0}

\section{Introduction}

The idea that our universe is in some sense mathematical goes back 
at least to the 
Pythagoreans, and has been extensively discussed in the literature 
(see, \eg, 
\cite{Dirac31,Wigner67,Suppes69,Zuse76,Rucker82,BarrowTOE,BarrowPi,DaviesGod,Jackiw94,Lloyd97,toe,Schmidhuber97,Ladyman98,multiverse,multiverse4wheeler,Schmidhuber00,Wolfram02,Cohen03,Tipler05,McCabe06a,McCabe06,StandishBook,Wilczek06,Wilczek07}).
Galileo Galilei stated that the Universe is a grand book
written in the language of mathematics, and Wigner reflected on the 
``unreasonable effectiveness of mathematics in the natural sciences'' \cite{Wigner67}.
In this essay, I will push this idea to its extreme and argue that our universe {\it is} mathematics in a well-defined sense.
After elaborating on this hypothesis and underlying assumptions in 
\Sec{HypothesisSec}, I discuss a variety of its 
implications in Sections~\ref{ScratchSec}-\ref{GoedelSec}. 
This paper can be thought of as the sequel to one I wrote in 1996 \cite{toe}, 
clarifying and extending the ideas described therein.

\section{The Mathematical Universe Hypothesis}
\label{HypothesisSec}

\subsection{The External Reality Hypothesis}

In this section, we will discuss the following two hypotheses and argue that, 
with a sufficiently broad definition of mathematical structure, the former implies the latter.

\smallskip
\centerline{\framebox{\parbox{7cm}{
{\bf External Reality Hypothesis (ERH):} {\it There exists an external
physical reality completely independent of us humans.}
}}}
\smallskip

\smallskip
\centerline{\framebox{\parbox{7cm}{
{\bf Mathematical Universe Hypothesis (MUH):} {\it Our external physical reality is a mathematical structure.}
}}}
\smallskip

\noindent
Although many physicists subscribe to the ERH and dedicate their careers to the 
search for a deeper understanding of this assumed external reality, 
the ERH is not universally accepted, and is rejected by, \eg, metaphysical solipsists.
Indeed, adherents of the Copenhagen interpretation of quantum mechanics may reject the ERH on the grounds that 
there is no reality without observation.
In this paper, we will assume that the ERH is correct and explore its implications.
We will see that, although it sounds innocuous, 
the ERH has sweeping implications for physics if taken seriously.

Physics theories aim to describe how this assumed external reality works.
Our most successful physics theories to date are generally regarded as 
descriptions of merely limited aspects of the external reality.
In contrast, the holy grail of theoretical physics is to find a {\it complete} description of it, 
jocularly referred to as a ``Theory of Everything'', or ``TOE''.

The ERH implies that for a description to be complete, it must be well-defined also according to 
non-human sentient entities (say aliens or future supercomputers) that lack the common understanding 
of concepts that we humans have evolved, \eg, ``particle'', ``observation'' or indeed any other 
English words.
Put differently,
such a description must be expressible 
in a form that is devoid of human ``baggage''.

\begin{figure}[pbt]\
\centerline{{\vbox{\epsfxsize=9.0cm\epsfbox{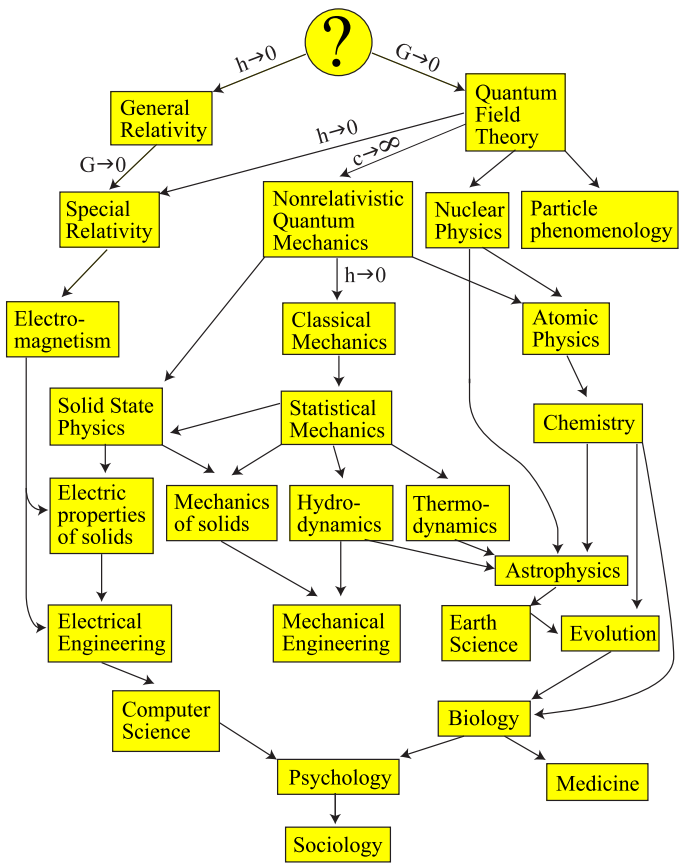}}}}
\caption{
Theories can be crudely organized into a family tree where
each might, at least in principle, be derivable
from more fundamental ones above it.
For example, classical mechanics can be obtained from 
special relativity in the approximation that the speed of light 
$c$ is infinite, and hydrodynamics with its concepts such as 
density and pressure can be derived from statistical 
mechanics. However, these cases where the arrows are well understood
form a minority.
Deriving biology from chemistry 
or psychology from biology appears unfeasible
in practice.
Only limited and approximate aspects of such subjects are mathematical, and it is likely that all mathematical 
models found in physics so far are similarly approximations of limited aspects of reality.
}
\label{TheoryTreeFig}
\end{figure}

\subsection{Reducing the baggage allowance}
\label{BaggageSec}

To give a few examples, \fig{TheoryTreeFig} illustrates how various  
theories can be crudely organized in a family tree where
each might, at least in principle, be derivable
from more fundamental ones above it.
All these theories have two components: mathematical equations and ``baggage'', 
words that explain how they are connected to what we humans observe and intuitively understand. 
Quantum mechanics as usually presented in textbooks has both 
components: some equations as well as three fundamental postulates
written out in plain English.
At each level in the hierarchy of theories, new concepts 
(\eg, protons, atoms, cells, organisms, cultures) are introduced 
because they are convenient, capturing the essence of what
is going on without recourse to the more fundamental theory
above it.  
It is important to remember, however, 
that it is we humans who introduce these concepts and the words
for them: in principle, everything could have been derived
from the fundamental theory at the top of the tree, although 
such an extreme reductionist approach appears
useless in practice.
Crudely speaking, the ratio of equations to
baggage decreases as we move down the tree, dropping near zero for 
highly applied fields such as medicine and sociology.
In contrast, theories near the top are highly mathematical, and 
physicists are still struggling to articulate the concepts, if any, 
in terms of which we can understand them. The MUH implies that the TOE indicated by the question mark 
at the top is purely mathematical, with no baggage whatsoever.

As an extreme example of a ``theory'', the description of external reality found in Norse mythology 
involves a gigantic tree named Yggdrasil, whose trunk supports Earth. This description all on its own 
is 100\% 
baggage, since it lacks definitions of ``tree'', ``Earth'', etc.
Today, the baggage fraction of this theory could be reduced by describing a tree as a particular arrangement of 
atoms, and describing this in turn as a particular quantum field theory state.
Moreover (see, \eg, \cite{Wigner67}), physics has come to focus on the way the external reality {\it works} 
(described by regularities known as laws of physics)
rather than on the way it {\it is} (the subject of initial conditions).

However, could it ever be possible to give a description of the external reality involving {\it no} baggage?
If so, our description of entities in the external reality and relations between them would have to be completely abstract, 
forcing any words or other symbols used to denote them to be mere labels with no preconceived meanings whatsoever.

A {\it mathematical structure} is precisely this: {\it abstract entities with relations between them}. 
Familiar examples include the integers and the real numbers.
We review detailed definitions of this and related mathematical notions in Appendix A. Here, let us instead illustrate 
this idea of baggage-free description with simple examples.
Consider the mathematical structure known as the group with two elements, \ie, addition modulo two.
It involves two elements that we can label ``0'' and ``1'' satisfying the following relations:
\beq{sumC2defEq}
\left\{
\begin{tabular}{l}
$0+0=0$\\
$0+1=1$\\
$1+0=1$\\
$1+1=0$
\end{tabular}
\right.
\eeq
The two alternative descriptions 
\beq{productC2defEq}
\left\{
\begin{tabular}{l}
$e\times e=e$\\
$e\times a=a$\\
$a\times e=a$\\
$a\times a=e$
\end{tabular}
\right.
\eeq
and
\beq{VerbalC2defEq}
\left\{
\begin{tabular}{l}
Even and even make even\\
Even and odd make odd\\
Odd and even make odd\\
Odd and odd make even
\end{tabular}
\right.
\eeq
look different, but the identification 
``0''$\to$``$e$''$\to$``even'', 
``1''$\to$``$a$''$\to$``odd'', 
``$+$''$\to$``$\times$''$\to$``and'', 
``$=$''$\to$``$=$''$\to$``make'' shows that they
describe exactly the same mathematical structure, 
since the symbols themselves are mere labels without intrinsic meaning.
The only intrinsic properties of the entities are those embodied by the relations between them.

\Eq{productC2defEq} suggests specifying the relations more compactly as a multiplication table.
Alternatively, using the notation of \eq{sumC2defEq}, a tabulation of the values of the function ``$+$"
for all combinations of the two arguments reads
\beq{C2StructureEq}
\begin{tabular}{|cc|}
\hline
$0$&$1$\\
$1$&$0$\\
\hline
\end{tabular}.
\eeq

In Appendix A, we give our convention
for encoding {\it any} finite mathematical structure, 
involving arbitrarily many entities, entity types and relations, as a finite sequence of integers.
Our present example corresponds to ``11220000110'', and Table~{\StructureTable}
lists a few other simple examples. Appendix A also covers infinite mathematical structures 
more relevant to physics, \eg, vector fields and Lie Groups, and how each of these can be encoded as a 
finite-length bit string.
In all cases, there are many equivalent ways of describing the same structure, and a particular mathematical 
structure can be defined as an {\it equivalence class of descriptions}. 
Thus although any one description involves some degree of arbitrariness (in notation, \etc), 
there is nothing arbitrary about the mathematical structure itself.

\begin{table}
\noindent
{\footnotesize {\bf Table~\StructureTable} -- Any finite mathematical structure can be encoded as 
a finite string of integers, which can in turn be encoded as a single bit string or integer.
The same applies to computable infinite mathematical structures as discussed in Appendix A.
\smallskip
}
\begin{tabular}{|l|l|}
\hline
Mathematical structure	&Encoding\\
\hline
The empty set		&100\\
The set of 5 elements 	&105\\
The trivial group $C_1$	&11120000\\
The polygon $P_3$	&113100120\\
The group $C_2$		&11220000110\\
Boolean algebra		&11220001110\\
The group $C_3$		&1132000012120201\\
\hline
\end{tabular}
\end{table}

Putting it differently, the trick is to ``mod out the baggage'', defining 
a mathematical structure by a description thereof modulo any freedom in notation. 
Analogously, the number 4 is well-defined even though 
we humans have multiple ways of referring to it, like ``IV'', ``four", ``cuatro'', ``fyra", \etc
 
\subsection{Implications for a Mathematical Universe}

In summary, there are two key points to take away from our discussion above and in Appendix A:
\begin{enumerate}
\item The ERH implies that a 
``theory of everything'' has no baggage.
\item Something that has a baggage-free description is precisely a mathematical structure. 
\end{enumerate}
Taken together, this implies the Mathematical Universe Hypothesis formulated 
on the first page of this article, \ie, 
that the external physical reality described by the TOE is a mathematical 
structure.\footnote{In the philosophy literature, the name
``structural realism'' has been coined for the doctrine that
the physical domain of a true theory corresponds to a mathematical structure \cite{Suppes69},
and the name ``universal structural realism'' has been used for 
the hypothesis that the physical universe is isomorphic to a mathematical structure \cite{McCabe06}.
}%\footnote{%

Before elaborating on this, let us consider a few historical examples to illustrate what it means
for the physical world to be a mathematical structure:
\begin{enumerate}
\item {\bf Newtonian Gravity of point particles:} Curves in $\R^4$ minimizing the Newtonian action.
\item {\bf General Relativity:} A 3+1-dimensional pseudo-Riemannian manifold with tensor fields obeying partial 
differential equations of, say, Einstein-Maxwell-theory with a perfect fluid component.
\item {\bf Quantum Field Theory:} Operator-valued fields on $\R^4$ obeying certain Lorentz-invariant partial differential equations 
and commutation relationships, acting on an abstract Hilbert space.
\end{enumerate}
(Of course, the true mathematical structure isomorphic to our world, if it exists, has not yet been found.)

When considering such examples, we need to distinguish between two 
different ways of viewing the external physical reality: 
the outside view or {\it bird perspective} of a mathematician 
studying the mathematical structure
and the inside view or {\it frog perspective} of an observer living 
in it.\footnote{Related ideas on ``observation from inside/outside the universe'' (albeit in a less mathematical context)
trace back to Archimedes and the 18th century physicist B.~J.~Boskovic --- see \cite{Roesssler87,Svozil94,Svozil96}.
}

A first subtlety in relating the two perspectives involves time. 
Recall that a mathematical structure is an abstract, immutable entity existing outside of space and time. If history
were a movie, the structure would therefore correspond not to a single frame of it but to the entire videotape.
Consider the first example above, a world made up of classical point particles moving around in three-dimensional Euclidean space
under the influence of Newtonian gravity.
In the four-dimensional spacetime of the bird perspective, these particle trajectories resemble a tangle of spaghetti. If
the frog sees a particle moving with constant velocity, the bird sees a straight strand of uncooked spaghetti. If the
frog sees a pair of orbiting particles, the bird sees two spaghetti strands intertwined like a double helix. To the
frog, the world is described by Newton's laws of motion and gravitation. To the bird, it is described by the geometry
of the pasta, obeying the mathematical relations corresponding to minimizing the Newtonian action.

A second subtlety in relating the two perspectives involves the observer.
In this Newtonian example, the frog (a self-aware substructure, or ``SAS'') itself must be merely a thick bundle of pasta, 
whose highly complex intertwining corresponds to a cluster of particles that store and process information.  
In the General Relativity example above, the frog 
is a tube through spacetime, a thick version of what Einstein referred to as a world-line.
The location of the tube would specify its position in space at different times.
Within the tube, the fields would exhibit complex behavior corresponding
to storing and processing information about the field-values in the surroundings,
and at each position along the tube, these processes would give rise to 
the familiar sensation of self-awareness.
The frog would perceive this one-dimensional 
string of perceptions along the tube as passage of time.
In the Quantum Field Theory example above, things become more subtle because 
a well-defined state of an observer can evolve into a quantum superposition of subjectively different states.
If the bird sees such deterministic frog branching, the frog perceives apparent randomness \cite{Everett57,EverettBook}.
Fundamental randomness from the bird's view is by definition banished (\Sec{RandomnessSec})

\subsection{Description versus equivalence}
\label{DescriptionSec}

Let us clarify some nomenclature.
Whereas the customary terminology in physics textbooks is that 
the external reality is {\it described by} mathematics, the MUH states that it {\it is} mathematics
(more specifically, a mathematical structure).
This corresponds to the ``ontic'' version of universal structural realism in the philosophical 
terminology of \cite{Ladyman98,McCabe06}.
If a future physics textbook contains the TOE, then its equations 
are the complete description of the mathematical structure that is the external physical reality.
We write {\it is} rather than {\it corresponds to} here, because if
two structures are isomorphic, then there is 
no meaningful sense in which they are not one and the same \cite{Cohen03}.  
From the definition of a mathematical structure (see Appendix  A), it follows that if
there is an isomorphism between a mathematical structure and another structure (a one-to-one correspondence 
between the two that respects the relations), then they are one and the same.
If our external physical reality is isomorphic to a mathematical structure, it therefore
fits the definition of being a mathematical structure.

If one rejects the ERH, one could argue that our universe is somehow made of stuff perfectly described
by a mathematical structure, but which also has other properties that are not described by it, and cannot be
described in an abstract baggage-free way.
This viewpoint, corresponding to the ``epistemic'' version of universal structural realism in the philosophical 
terminology of \cite{Ladyman98,McCabe06},
would make Karl Popper turn in his grave, since those additional bells
and whistles that make the universe non-mathematical by definition have no observable effects whatsoever.

\subsection{Evidence for the MUH}

Above we argued that the ERH implies the MUH, so that any evidence for the ERH is also evidence for the MUH.
Before turning to implications of the MUH, let us briefly discuss additional reasons for taking 
this hypothesis seriously.

In his above-mentioned 1967 essay \cite{Wigner67}, Wigner argued that 
``the enormous usefulness of mathematics in the
natural sciences is something bordering on the mysterious", and that
``there is no rational explanation for it".
The MUH provides this missing explanation.
It explains the utility of mathematics for describing the physical
world as a natural consequence of the fact that the latter {\it is}
a mathematical structure, and we are simply uncovering this 
bit by bit.
The various approximations that constitute our current physics theories
are successful because simple mathematical
structures can provide good approximations
of certain aspects of more complex mathematical structures.
In other words, our successful theories are
not mathematics approximating physics,
but mathematics approximating mathematics.

The MUH makes the testable prediction that further mathematical regularities 
remain to be uncovered in nature.
This predictive power of the mathematical universe idea was expressed by Dirac in 1931:
``The most powerful method of advance that can be suggested at present is to
employ all the resources of pure mathematics in attempts to perfect and
generalize the mathematical formalism that forms the existing basis of
theoretical physics, and after each success in this direction, to try to
interpret the new mathematical features in terms of physical entities'' \cite{Dirac31}.
After Galileo promulgated the mathematical universe idea, 
additional mathematical regularities beyond his wildest dreams were
uncovered, ranging from the motions of planets to the properties of atoms.
After Wigner had written his 1967 essay \cite{Wigner67},  
the standard model of particle physics 
revealed new ``unreasonable'' mathematical 
order in the microcosm of elementary particles and in the macrocosm of the early universe.
I know of no other compelling explanation for this trend
than that the physical world really is completely mathematical.

\subsection{Implications for Physics}
\label{ImplicationsSubSec}

The notion that the world is a mathematical structure 
alters the way we view 
many core notions in physics. In the remainder of this paper, 
we will discuss numerous such examples, ranging from standard topics
like symmetries, irreducible representations, units, free parameters and initial conditions 
to broader issues like parallel universes, G\"odel incompleteness and the idea of our reality being a computer simulation.

\section{Physics from scratch}
\label{ScratchSec}

Suppose we were given mathematical equations that completely describe the physical world, including us, 
but with no hints about how to interpret them. What would we do with them?\footnote{The 
reader may prefer a TOE that includes such human-language hints even if they are in principle redundant.
Indeed, the conventional approach in the philosophy of science holds that a 
theory of mathematical physics can be broken down into (i) a mathematical structure, (ii) an empirical domain 
and (iii) a set of correspondence rules which link parts of the mathematical structure with parts of the empirical 
domain. 
In our approach, (i) is the bird's view, (ii) is the frog's view, and the topic of the present section is how to derive both
(ii) and (iii) from (i) alone.
Analogously,
given an abstract but complete description of a car (essentially the locations of its atoms),
someone wanting practical use of this car may be more interested in a hands-on user's manual 
describing which are the most important components and how to use them.
However, by carefully examining the original description or the car itself, 
she might be able to figure out how the car works and write her own manual.
}

Specifically, what mathematical analysis of them would reveal their phenomenology, \ie, the
properties of the world that they describe as perceived by observers?
Rephrasing the question in the terminology of the previous section: given a mathematical structure, 
how do we compute the inside view from the outside view?
More precisely, we wish to derive the 
``consensus view''\footnote{As discussed in \cite{multiverse4wheeler},
the standard mental picture of what the physical world is corresponds  
to a third intermediate viewpoint that could be termed the 
{\it consensus view}.
From your subjectively perceived frog perspective, the world turns upside down when you stand on your head
and disappears when you close your eyes, yet you subconsciously interpret your sensory inputs 
as though there is an external reality that is 
independent of your orientation, your location and your state of mind.
It is striking that although this third view involves both
censorship (like rejecting dreams),
interpolation (as between eye-blinks)
and extrapolation (say attributing existence to unseen cities)
of your inside view, independent observers nonetheless appear to share this 
consensus view.
Although the inside view looks black-and-white to a cat, 
iridescent to a bird seeing four primary colors,
and still more different to a bee seeing polarized light, a bat using sonar,
a blind person with keener touch and hearing, or the latest 
robotic vacuum cleaner, all agree on whether the door is open.
The key current challenge in physics is deriving this semiclassical 
consensus view from the fundamental equations specifying the bird perspective.
In my opinion, this means that 
although understanding the detailed nature of human consciousness is an
important challenge in its own right, it is {\it not} necessary 
for a fundamental theory of physics, 
which, in the case of us humans, corresponds to the mathematical description of our world found in physics textbooks.
It is not premature to address this question now, before we have found such equations, since it is
important for the search itself --- without answering it, we will not know if a given candidate theory is consistent with what we observe.
}.

This is in my opinion one of the most important questions facing theoretical physics, because if we cannot 
answer it, then we cannot test a candidate TOE by confronting it with observation.
I certainly do not purport to have a complete answer to this question. This subsection merely explores some possibly useful first 
steps to take in addressing it.

By construction, the only tools at our disposal are purely mathematical ones, so the only way in which 
familiar physical notions and interpretations (``baggage'') can emerge are as implicit properties of the structure itself that 
reveal themselves during the mathematical investigation. What such mathematical investigation is relevant?
The detailed answer clearly depends on the nature of the mathematical structure being investigated, but a useful 
first step for {\it any} mathematical structure $S$ is likely to be finding its {\it automorphism group} $\Aut(S)$,
which encodes its symmetries.

\subsection{Automorphism definition and simple examples} 

A mathematical structure $S$ (defined and illustrated with examples in Appendix A) is essentially
a collection of abstract entities with relations (functions) between them \cite{Hodges}.
We label the various sets of entities $S_1$, $S_2$, ... and the functions 
(which we also refer to as {\it relations}\footnote{As discussed in Appendix A, this is a slight 
generalization of the customary notion of a relation, which corresponds to the special case of a function 
mapping into the two-element (Boolean) set $\{0,1\}$ or $\{False,True\}$.}) $R_1$, $R_2$, .... 
An {\it automorphism} of a mathematical structure $S$ is defined as a permutation $\sigma$ of the
elements of $S$ that preserves all relations. %(all functions, as defined in \Sec{StructureDefSec}).
Specifically, 
$R_i = R_i'$ for all functions with 0 arguments, 
$R(a') = R(a)'$ for all functions with 1 argument,
$R(a',b') = R(a,b)'$ for all functions with 2 argument, \etc, 
where $'$ denotes the action of the permutation. 
It is easy to see that the set of all automorphisms of a structure, denoted $\Aut(S)$, form a group. 
$\Aut(S)$ can be thought of as the group of symmetries of the structure, \ie, the transformations under which the structure is invariant.
Let us illustrate this with a few examples.

The trivial permutation $\sigma_0$ (permuting nothing) is of course an automorphism, so $\sigma_0\in\Aut(S)$ for any structure.
Mathematical structures which have no non-trivial automorphisms (\ie, $\Aut(S)=\{\sigma_0$\}) are called {\it rigid} and exhibit no symmetries.
Examples of rigid structures include Boolean algebra (\Sec{BooleanExampleSec}),
the integers and the real numbers.
At the opposite extreme, consider a set with $n$ elements and no relations. All $n!$ permutations of this mathematical structure
are automorphisms; there is total symmetry and no element has any features that distinguishes it from any other element.

For the simple example of the 3-element group defined by \eq{Z3StructureEq},
there is only one non-trivial automorphism $\sigma_1$, corresponding to swapping the two non-identity elements:
$e'=e$, $g_1'=g_2$, $g_2'=g_1$, so $\Aut(S)=\{\sigma_0,\sigma_1\}=C_2$, the group of two elements.

\subsection{Symmetries, units and dimensionless numbers}
\label{SumUnitDimlessSec}

For a more physics-related example, consider the mathematical structure of 3D Euclidean space defined as 
a vector space with an inner product as
follows:\footnote{A more careful definition would circumvent the issue that 
$R_2(x_0,x_0)$ is undefined. We 
return to the issue of G\"odel-completeness and computability for continuous structures in 
\Sec{GoedelSec}.}
\begin{itemize}
\itemsep0cm
\item $S_1$ is a set of elements $x_\alpha$ labeled by real numbers $\alpha$,
\item $S_2$ is a set of elements $y_\rvec$ labeled by 3-vectors $\rvec$, % (ordered triplets of real numbers) 
\item $R_1(x_{\alpha_1},x_{\alpha_2})=x_{\alpha_1-\alpha_2}\in S_1$,
\item $R_2(x_{\alpha_1},x_{\alpha_2})=x_{\alpha_1/\alpha_2}\in S_1$,
\item $R_3(y_{\rvec_1},y_{\rvec_2}) = y_{\rvec_1+\rvec_2}\in S_2$, 
\item $R_4(x_\alpha,y_\rvec) = y_{\alpha\rvec}\in S_2$,
\item $R_5(y_{\rvec_1},y_{\rvec_2}) = x_{\rvec_1\cdot\rvec_2}\in S_1$.
\end{itemize}
Thus we can interpret $S_1$ as the (rigid) field of real numbers and $S_2$ as 3D Euclidean space,
specifically the vector space $\Reals^3$ with Euclidean inner product.
By combining the relations above, all other familiar relations can be generated, for example 
the origin $R_1(x_\alpha,x_\alpha)=x_0$ and multiplicative identity
$R_2(x_\alpha,x_\alpha)=x_1$ in $S_1$ together with the 
additive inverse $R_1(R_1(x_\alpha,x_\alpha),x_\alpha)=x_{-\alpha}$ and the 
multiplicative inverse $R_2(R_2(x_\alpha,x_\alpha),x_\alpha)=x_{\alpha^{-1}}$,
as well as the 3D origin $R_4(R_1(x_\alpha,x_\alpha),y_\rvec)=y_{\zero}$.
This mathematical structure has rotational symmetry, \ie, the automorphism group $\Aut(S)=O(3)$, 
parametrized by $3\times 3$ rotation matrices $\R$ acting as follows:
\beqa{NormedVectorAutomorphismEq1}
x_\alpha'	&=&x_\alpha,\\
y_\rvec'  	&=&y_{\R\rvec}.
\eeqa
To prove this, one simply needs to show that each generating relation respects the symmetry:
\beqa{NormedVectorAutProofEq1}
R_3(y_{\rvec_1}',y_{\rvec_2}')&=&y_{\R\rvec_1+\R\rvec_2} =  y_{\R(\rvec_1+\rvec_2)} = R_3(y_{\rvec_1},y_{\rvec_2})',\nonumber\\
R_4(x_\alpha',y_\rvec')&=&y_{\alpha\R\rvec} =  y_{\R\alpha\rvec}=R_4(x_\alpha,y_\rvec)',\nonumber\\
R_5(y_{\rvec_1}',y_{\rvec_2}')&=&x_{(\R\rvec_1)\cdot(\R\rvec_2)}=x_{\rvec_1\cdot\rvec_2}=R_5(y_{\rvec_1},y_{\rvec_2})'.
\nonumber
\eeqa
\etc

This simple example illustrates a number of points relevant to physics.
First of all, the MUH implies that {\it any symmetries in the mathematical structure correspond to physical symmetries},
since the relations $R_1$, $R_2$, ... exhibit these symmetries and these relations are the 
{\it only} properties that the set elements have. An observer in a space defined as $S_2$ above 
could therefore not tell the difference between this space and a rotated version of it.

Second, {\it relations are potentially observable}, because they are properties of the structure.
It is therefore crucial to define mathematical structures precisely, since 
seemingly subtle differences in the definition can make a crucial difference for the physics.
For example, the manifold $\Reals$, the metric space $\Reals$, the vector space $\Reals$ and the number field $\Reals$ are all 
casually referred to as simply ``$\Reals$'' or ``the reals'', yet they are four different structures with 
four very different symmetry groups.
Let us illustrate this with examples related to the mathematical description of our 3-dimensional physical space, and see how
such considerations can rule out many mathematical structures as candidates for corresponding to our universe.

The 3D space above contains a special point, its origin $y_{\zero}$
defined by 
$R_4(R_1(x_\alpha,x_\alpha),y_\rvec)$ above, with no apparent counterpart in our physical space.
Rather (ignoring spatial curvature for now), our physical space appears to have a further symmetry, translational symmetry, which
this mathematical structure lacks.
The space defined above thus has too much structure.
This can be remedied by dropping $R_3$ and $R_4$ 
and replacing the 5th relation by 
\beq{MetricSpaceEq}
R_5(y_{\rvec_1},y_{\rvec_2})=x_{|\rvec_1-\rvec_2|}, 
\eeq
thereby making $S_2$ a metric 
space rather than a vector space.

However, this still exhibits more structure than our physical space: it has a preferred length scale. Herman Weyl 
emphasized this point in \cite{Weyl22}.
Lengths are measured by real numbers from $S_1$ which form a rigid structure, where the 
multiplicative identity $x_1$ is special and singled out
by the relation $R_2(x_\alpha,x_\alpha)=x_1$. In contrast, there appears to be no length scale ``1'' of special significance in our physical space.
Because they are real numbers,
two lengths in the mathematical structure can be multiplied to give another length.
In contrast, we assign different units to length and area in our physical space because they cannot be directly compared.
The general implication is that {\it quantities with units are not real numbers}. 
Only dimensionless quantities in physics may correspond to real numbers in the mathematical structure.
Quantities with units may instead correspond to the 1-dimensional vector space over the reals,
so that only ratios between quantities are real numbers.
The simplest mathematical structure corresponding to the Euclidean space of classical physics (ignoring relativity)
thus involves three sets:
\begin{itemize}
\itemsep0cm
\item $S_1$ is a set of elements $x_\alpha$ labeled by real numbers $\alpha$,
\item $S_2$ is a set of elements $y_\alpha$ labeled by real numbers $\alpha$,
\item $S_3$ is a set of elements $z_\rvec$ labeled by 3-vectors $\rvec$,
\item $R_1(x_{\alpha_1},x_{\alpha_2})=x_{\alpha_1-\alpha_2}\in S_1$,
\item $R_2(x_{\alpha_1},x_{\alpha_2})=x_{\alpha_1/\alpha_2}\in S_1$,
\item $R_3(y_{\alpha_1},y_{\alpha_2)}=y_{\alpha_1+\alpha_2}\in S_2$,
\item $R_4(x_{\alpha_1},y_{\alpha_2}) = y_{\alpha_1\alpha_2}\in S_2$,
\item $R_5(z_{\rvec_1},z_{\rvec_2},z_{\rvec_3}) = y_{(\rvec_2-\rvec_1)\cdot(\rvec_3-\rvec_1)}\in S_2$.
\end{itemize}
Here $S_1$ is the rigid field of real numbers $\Reals$, $S_2$ is the 1-dimensional vector space $\Reals$ (with no division and 
no preferred length scale) and $S_3$ is a metric space (corresponding to physical space) 
where angles are defined but lengths are only defined up to an overall scaling.
In other words, three points define an angle, and two points define a distance
via the relation 
$R_5(z_{\rvec_1},z_{\rvec_2},z_{\rvec_2}) = y_{|\rvec_2-\rvec_1|^2} \in S_2$.

The most general automorphism of this structure corresponds to 
(proper or improper) rotation by a matrix $\R$, translation by a vector $\a$ and scaling by a nonzero constant $\lambda$:
\beqa{EucAutomorphismEq1}
x_\alpha'	&=&x_\alpha,\nonumber\\
y_\alpha'  	&=&y_{\lambda^2\alpha},\nonumber\\
z_\rvec'  	&=&z_{\lambda\R\rvec+\a}.\nonumber
\eeqa
For example,
\beqa{EucAutomorphismEqProof}
&&R_5(z_{\rvec_1}',z_{\rvec_2}',z_{\rvec_3}')=\nonumber\\
&&=y_{[(\lambda\R\rvec_2+\a) - (\lambda\R\rvec_1+\a)]\cdot[(\lambda\R\rvec_3+\a) - (\lambda\R\rvec_1+\a)]}\nonumber\\
&&=y_{\lambda^2(\rvec_2-\rvec_1)\cdot(\rvec_3-\rvec_1)}=R_5(z_{\rvec_1},z_{\rvec_2},z_{\rvec_3})'.\nonumber\\
\nonumber
\eeqa

The success of general relativity suggests that our physical space possesses still more symmetry %(diffeomorphism symmetry),
whereby also relations between widely separated points (like $R_5$) are banished.
Instead, distances are defined only between infinitesimally close points.
Note, however, that neither diffeomorphism symmetry nor gauge symmetry correspond to automorphisms of the mathematical structure. 
In this sense, these are not physical symmetries, and correspond instead to redundant notation,  
\ie, notation transformations relating different equivalent descriptions of 
the same structure. %; see Appendix A for explicit examples illustrating this.
For recent reviews of these subtle issues and related controversies, see \cite{BrownBrading02,BrownCastellani03}.

If it were not for quantum physics, the mathematical structure of general relativity 
(a 3+1-dimensional pseudo-Riemannian manifold with various ``matter'' tensor fields obeying certain partial 
differential equations) would be a good candidate for the mathematical structure corresponding to our universe.
A fully rigorous definition of the mathematical structure corresponding to the $SU(3)\times SU(2)\times U(1)$ quantum field theory 
of the standard model is still considered an open problem in axiomatic field theory, even aside from the issue of quantum gravity.

\subsection{Orbits, subgroups and further steps}

Above we argued that when studying a mathematical structure $S$ to derive its physical phenomenology (the ``inside view''), a
useful first step is finding its symmetries, specifically its automorphism group $\Aut(S)$.
We will now see that this in turn naturally leads to further analysis steps, such as finding orbit partitions, irreducible actions 
and irreducible representations.

For starters, subjecting $\Aut(S)$ to the same analysis that mathematicians routinely perform when examining any group can reveal features 
with a physical flavor.
For example, computing the subgroups of $\Aut(S)$ for our last example reveals that there are subgroups 
of four qualitatively different types. We humans have indeed coined names (``baggage'') for them: translations, rotations, scalings and parity reversal.
Similarly, for a mathematical structure whose symmetries include the Poincar\'e group, 
translations, rotations, boosts, parity reversal and time reversal all emerge as separate notions in this way.

Moreover, the action of the group $\Aut(S)$ on the elements of $S$ partitions them into equivalence classes (known in group theory as ``orbits''), 
where the orbit of a given element is defined as the set of elements that $\Aut(S)$ can transform it into.
The orbits are therefore in principle observable and distinguishable from each other, 
whereas all elements on the same orbit are equivalent by symmetry.
In the last example, any 3D point in $S_3$ can be transformed into any other point in $S_3$, so all points in this space 
are equivalent. The scaling symmetry decomposes $S_2$ into two orbits ($0$ and the rest) 
whereas each element in $S_1$ is its own orbit and hence distinguishable. 
In the first example above, where $\Aut(S)$ is the rotation group, the three-dimensional points separate into distinct classes, 
since each spherical shell in $S_2$ of fixed radius is its own orbit.

\subsection{Group actions and irreducible representations}

Complementing this ``top-down'' approach of mathematically analyzing $S$, we can obtain further hints for useful mathematical 
approaches by observing our inside view of the world around us and investigating how this can be linked to the underlying 
mathematical structure implied by the MUH. We know empirically that our mathematical structure contains self-aware substructures (``observers'') able to
describe some aspects of their world mathematically, and that symmetry considerations play a major role in these mathematical descriptions.
Indeed, when asked ``What is a particle?'', many theoretical physicists like to smugly reply
``An irreducible representation of the Poincar\'e group''.
This refers to the famous insight by Wigner and others \cite{Majorana32,Dirac36,Proca36,Wigner39,Wigner67} that any mathematical 
property that we can assign to a quantum-mechanical
object must correspond to a ray representation of the group of spacetime symmetries.
Let us briefly review this argument and generalize it to our present context.

As discussed in detail in \cite{HoutappelvanDamWigner65,Wigner67}, the observed state of our physical world (``initial conditions'') 
typically exhibits no symmetry at all, whereas the perceived regularities (``laws of physics'') are invariant under some 
symmetry group $G$ (\eg, the Poincar\'e symmetry for the case of relativistic quantum field theory).  
Suppose we can describe some aspect of this state (some properties of a localized object, say) by a vector of numbers $\vpsi$.
Letting $\vpsi'=\rho_\sigma(\vpsi)$ denote the description after applying a symmetry transformation $\sigma\in G$, 
the transformation rule $\rho_\sigma$ must by definition 
have properties that we will refer to as identity, reflexivity and transitivity.
\begin{enumerate}
\item {\bf Identity:} $\rho_{\sigma_0}(\vpsi)=\vpsi$ for the identity transformation $\sigma_0$ 
(performing no transformation should not change $\vpsi$).
\item {\bf Reflexivity:} $\rho_{\sigma^{-1}}(\rho_\sigma(\vpsi))=\vpsi$ (transforming and then un-transforming should recover $\vpsi$).
\item {\bf Transitivity:} If $\sigma_2\sigma_1=\sigma_3$, then $\rho_{\sigma_2}(\rho_{\sigma_1}(\vpsi))=\rho_{\sigma_3}(\vpsi)$ 
(making two subsequent transformations should be equivalent to making the combined transformation).
\end{enumerate}
In other words, the mapping from permutations $\sigma$ to transformations $\rho_\sigma$ must be a homomorphism. % (a structure-preserving map).

Wigner and others studied the special case of quantum mechanics, where the description $\vpsi$ corresponded to a complex ray,
\ie, to an equivalence class of complex unit vectors where any two vectors were defined as equivalent if they only differed by an overall phase.
They realized that for this case, the transformation $\rho_\sigma(\vpsi)$ must be linear and indeed unitary, 
which means that it satisfies the definition of being a 
so-called ray representation (a regular unitary group representation up to a complex phase) of the symmetry group $G$.
Finding all such representations of the Poincar\'e group thus gave a catalog of all possible transformation properties that quantum objects could have
(a mass, a spin=0, 1/2, 1, ..., \etc),
essentially placing an upper bound on what could exist in a Poincar\'e-invariant world. 
This cataloging effort was dramatically simplified by
the fact that all representations can be decomposed into a simple list of irreducible ones, whereby degrees of freedom in $\vpsi$ can be partitioned
into disjoint groups that transform without mixing between the groups. This approach has proven useful not only for deepening our understanding 
of physics (and of how ``baggage'' such as mass and spin emerges from the mathematics), but also for simplifying many quantum calculations. 
Wigner and others have emphasized that to a large extent, 
``symmetries imply dynamics'' in the sense that dynamics is the transformation corresponding to time translation (one of the Poincar\'e symmetries), 
and this is in turn dictated by the irreducible representation.

For our case of an arbitrary mathematical structure, we must of course drop the quantum assumptions. 
The identity, reflexivity and transitivity properties alone then tell us merely that $\rho_\sigma$ is what 
mathematicians call a {\it group action} on the set of descriptions $\vpsi$. Here too, the notion of reducibility can be defined,
and an interesting open question is to what extent the above-mentioned representation theory results can be be generalized.
In particular, since the ray representations of quantum mechanics are more general than unitary representation and less general than group actions,
one may ask whether the quantum mechanics case is in some sense the only interesting group action, or whether there are others as well.
Such a group action classification could then be applied both to the exact symmetry group $\Aut(S)$ and to 
any effective or partial symmetry groups $G$, an issue to which we return below. 

\subsection{Angles, lengths, durations and probabilities}

Continuing this approach of including empirical observations as a guide, our discussion connects 
directly with results in the relativity and quantum literature.
Using the empirical observation that we can measure consensus view quantities we call angles, distances and durations,
we can connect these directly 
to properties of the Maxwell equations {\etc} in our mathematical structure. 
Pioneered by Einstein, this approach produced the theory of special relativity, complete with its transformation rules and
the ``baggage'' explaining in human language how these quantities were empirically measured. 
The same approach has been successfully pursued in the general relativity context, where it has proven both subtle and crucial for dealing with gauge ambiguities {\etc}
Even in minimal general relativity without electromagnetism, the above-mentioned 
empirical properties of space and time can be analogously derived, replacing light clocks by 
gravitational wave packets orbiting black holes, {\etc}

There is also a rich body of literature pursuing the analogous approach for quantum mechanics. 
Starting with the empirical fact that quantum observers perceive not 
schizophrenic mental superpositions but apparent randomness, the baggage
corresponding to the standard rule for computing the corresponding probabilities 
from the Hilbert space quantities has arguably been derived (see, \eg, \cite{Deutsch99} and 
references therein).

\subsection{Approximate symmetries}

Although the above analysis steps were all discussed in the context of the {\it exact} symmetries of the 
mathematical structure that by the MUH is our universe, they can also be applied to the {\it approximate} symmetries of 
mathematical structures that approximate certain aspects of this correct structure.
Indeed, in the currently popular view that all we have discovered so far are effective theories 
(see, \eg, \cite{WeinbergQFT97}),
all physical symmetries studied to date (except perhaps CPT symmetry) are likely to be of this approximate status.
Such effective structures can exhibit either more or less symmetry than the underlying one. 
For example, the distribution of air molecules around you exhibits no exact symmetry, but can be well approximated
by a continuous gas in thermal equilibrium obeying a translationally and rotationally invariant wave equation.
On the other hand, even if the quantum superposition of all field configurations emerging from cosmological inflation 
is a translationally and rotationally invariant state, any one generic element of this superposition, 
forming our classical ``initial conditions'', will lack this symmetry. We return to this issue in \Sec{ICsubsec}.

In addition, approximate symmetries have proven useful even when the approximation is rather inaccurate.
Consider, for example, a system governed by a Lagrangian, where one term exhibits a symmetry that is broken by another subdominant term.
This can give rise to a natural decomposition into subsystems and an emergence of attributes thereof.
For charged objects moving in an electromagnetic field, the energy-momentum 4-vectors of the objects would be
separately conserved (by Noether's theorem because these terms exhibit symmetry under time and space translation)
were it not for the object-field interaction terms in the Lagrangian,
so we perceive them as slowly changing property of the objects themselves. 
In addition, the full Lagrangian has spacetime translational symmetry, 
so total energy-momentum is strictly conserved when we include the energy-momentum 
of the electromagnetic field. These two facts together let us think of 
momentum as gradually flowing back and forth between the objects and the radiation.
More generally, partial symmetries in the Lagrangian provide a natural subsystem/subsector decomposition
with quantities that are conserved under some interactions but not others.
Quantities that are conserved under faster interactions (say the strong interaction) 
can be perceived as attributes of objects (say atomic number $Z$) that evolve because of 
a slower interaction (say the weak interaction).

\section{Implications for symmetry, initial conditions, randomness and physical constants}
\label{ICsec}

Above we explored the ``physics from scratch'' problem, \ie, possible approaches to 
deriving physics as we know it from an abstract baggage-free mathematical structure.
This has a number of implications for foundational physics questions related to 
symmetry, initial conditions and physical constants.

\subsection{Symmetry}

The way modern physics is usually presented, symmetries are treated as an input rather than
an output. For example, Einstein founded special relativity on 
Lorentz symmetry, \ie, on the
postulate that all laws of physics, including that governing the speed of light, are the same in 
all inertial frames. Likewise, the $SU(3)\times SU(2)\times U(1)$ symmetry of the standard
model is customarily taken as a starting assumption.

Under the MUH, the logic is reversed. The mathematical structure $S$ of our universe 
has a symmetry group $\Aut(S)$ that manifests itself in our perceived physical laws.
The laws of physics being invariant under $\Aut(S)$ (\eg, the Poincar\'e group) is therefore
not an input but rather a logical consequence of the way the inside view arises from the outside view.

Why do symmetries play such an important role in physics? 
The points by Wigner and others regarding their utility for calculations and insight 
were reviewed above. 
The deeper question of why our structure $S$ has so much symmetry is equivalent to the question of why we
find ourselves on this particular structure rather than another one with less symmetry.
It is arguably not surprising considering that symmetry 
(\ie, $\Aut(S)\ne\{\sigma_0$\}) 
appears to be more the rule than the exception in mathematical structures.
However, an anthropic selection effect may be at play as well: as pointed out by Wigner, the existence
of observers able to spot regularities in the world around them probably requires symmetries \cite{Wigner67}. 
For example,
imagine trying to make sense of a world where the outcome of experiments depended on
the spatial and temporal location of the experiment.

\begin{figure}[pbt]
\centerline{{\vbox{\epsfxsize=9.0cm\epsfbox{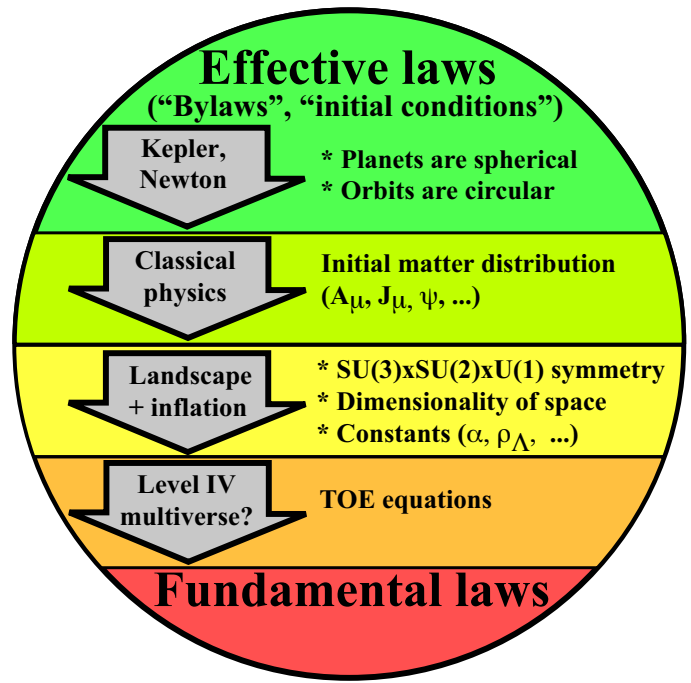}}}}

\caption{\label{BoundaryFig}
The shifting boundary (horizontal lines) 
between fundamental laws and environmental laws/effective laws/initial conditions.
Whereas Ptolemy and others had argued that the circularity of
planetary orbits was a fundamental law of nature, Kepler and Newton reclassified this as
an initial condition, showing that the fundamental laws also 
allowed highly non-circular orbits.
Classical physics removed from the fundamental law category also 
the initial conditions for the electromagnetic field and 
all other forms of matter and energy (responsible for almost all the complexity we observe), 
leaving the fundamental laws quite simple.
A TOE with a landscape and inflation reclassifies many of the remaining ``laws'' as initial conditions,
since they can differ from one post-inflationary region to another, but since 
inflation generically makes each such region infinite, it can 
fool us into misinterpreting these environmental properties as fundamental laws.
Finally, if the MUH is correct and the Level IV multiverse of all mathematical structures (see \Sec{PUsec}) exists, 
then even the ``theory of everything'' equations that physicists are seeking are merely local bylaws
in Rees' terminology\cite{ReesHabitat}, differing across the ensemble.
}
\end{figure}

\subsection{Initial conditions}
\label{ICsubsec}

The MUH profoundly affects many notions related to initial conditions.
The traditional view of these matters is eloquently summarized by {\eg} \cite{HoutappelvanDamWigner65,Wigner67} as 
splitting our quantitative description of the world into two domains, ``laws of physics'' and ``initial conditions''.
The former we understand and hail as the purview of physics, 
the latter we lack understanding of and merely take as an input to our calculations.

\subsubsection{How the MUH banishes them}
 
As illustrated in \fig{BoundaryFig}, the borderline between these two domains has gradually shifted at the expense of initial conditions.
Newton's orbital calculations focused on how our solar system evolved, not on how it came into existence, yet solar system formation is now 
a mainstream research area. Gradually pushing the frontier of our ignorance further back in time, scientists have studied the formation of galaxies
billions of years ago, 
the synthesis of atomic nuclei during the first few minutes after our Big Bang and the formation of density fluctuations during an inflationary
epoch many orders of magnitude earlier still.
Moreover, a common feature of much string theory related model building is that 
there is a ``landscape'' of solutions, corresponding to spacetime configurations involving 
different dimensionality, different types of fundamental particles and different values for certain physical ``constants''
(see Table 1 in \cite{axions} for an up-to-date list of the 32 parameters specifying the standard models of particle physics and cosmology),
some or all of which may vary across the landscape. 
Eternal inflation transforms such potentiality into reality, actually creating regions of space realizing each of
these possibilities. However, each such region where inflation has ended is generically infinite in size, potentially making 
it impossible for any inhabitants to travel to other regions where these apparent laws of physics are different.
If the MUH is correct and the Level IV multiverse of all mathematical structures (see \Sec{PUsec}) exists, 
this historical trend is completed: even the ``theory of everything'' equations that physicists are seeking are an environmental accident,
telling us not something fundamental about reality, but instead which particular mathematical structure we happen to inhabit, like a 
multiversal telephone number.

In other words, this would entail a crushing complete defeat of fundamental physical laws. 
However, contrary to how it may at first appear, it would not constitute a victory for initial 
conditions in the traditional sense. There is nothing ``initial'' about specifying a mathematical structure.
Whereas the traditional notion of initial conditions entails that our universe ``started out'' in some particular state,
mathematical structures do not exist in an external space or time, are not created or destroyed, and in many cases also lack any internal 
structure resembling time.
Instead, {\it the MUH leaves no room for ``initial conditions''}, eliminating them altogether.
This is because the mathematical structure is by definition a {\it complete} description of the physical world.
In contrast, a TOE  
saying that our universe just ``started out'' or ``was created'' in some unspecified state constitutes
an incomplete description, thus violating both the MUH and the ERH.

\subsubsection{How they are a useful approximation}

We humans may of course be unable to measure certain properties of our world and unable to determine which
mathematical structure we inhabit. Such epistemological uncertainty is clearly compatible with the MUH, and
also makes the notion of initial conditions a useful approximation even if it lacks a fundamental basis.
We will therefore continue using the term initial conditions in this limited sense throughout this paper.

To deal with such uncertainty in a quantitative way, we humans have invented statistical mechanics 
and more general statistical techniques for quantifying statistical relations between observable 
quantities. These involve the notion that our initial conditions are a member of a 
real or hypothetical ensemble of possible initial conditions, and quantify how atypical our
initial conditions are by their entropy or algorithmic complexity 
(see, \eg, \cite{Chaitin90,LiVitanyiBook}).

\subsubsection{How the MUH banishes randomness}
\label{RandomnessSec}

By insisting on a complete description of reality, the MUH banishes not only the classical notion of 
initial conditions, but also the classical notion of randomness.
The traditional view of randomness (viewed either classically or as in the Copenhagen interpretation of quantum mechanics)
is only meaningful in the context of an external time, 
so that one can start with one state and then have something random ``happen'', causing 
two or more possible outcomes.
In contrast, the only intrinsic properties of a mathematical structure are its relations,
timeless and unchanging.
In a fundamental sense, the MUH thus implies Einstein's dictum ``God does not play dice''.

This means that if the MUH is correct, the only way that randomness and probabilities can appear in physics 
is via the presence of ensembles, as a way for observers to quantify their ignorance about which element(s) of the ensemble they are in. 
Specifically, all mathematical statements about probability can be recast as measure theory. 
For example, if an observer has used a symmetric quantum random number generator to produce a bit string written out as a real number 
like ".011011011101...'', and if quantum mechanics is unitary so that the final state is a superposition of observers obtaining all outcomes,
then in the limit of infinitely many bits, almost all observers will find their bit strings to appear perfectly random and conclude
that the conventional quantum probability rules hold.
This is because according to Borel's theorem on normal numbers \cite{Borel,Chung},
almost all (all except for a set of Borel measure zero) real numbers have binary decimals passing the standard tests of 
randomness.\footnote{It is interesting to note that Borel's 1909 theorem 
made a strong impression on many mathematicians of the time, some of whom
had viewed the entire probability concept with a certain suspicion,
since they were now confronted with a 
theorem in the heart of classical mathematics
which could be reinterpreted in terms of probabilities \cite{Chung}.
Borel would undoubtedly have been interested to know that his work demonstrated 
the emergence of a probability-like concept ``out of the blue'' 
not only in mathematics, but in physics as well.
}
A convincing demonstration that there is such a thing as true randomness in the laws of physics
(as opposed to mere ensembles where epistemological uncertainty grows) would therefore refute the MUH.

\subsubsection{Cosmic complexity}
\label{ComplexitySec}

There is an active literature on the complexity of our observable
universe and how natural it is (see, \eg, \cite{Davies,nihilo,ZehTimeBook,AlbrechtSorbo04,CarrollChen05,Wald05,Page06,Vilenkin07} and references therein), 
tracing back to Boltzmann \cite{Boltzmann} and others. 
The current consensus is that the initial conditions (shortly before Big Bang nucleosynthesis, say) of this comoving volume of space (our so-called Hubble volume)
were both enormously complex and yet surprisingly simple.
To specify the state of its $\sim 10^{78}$ massive particles (either classically in terms of positions and velocities, or quantum-mechanically)
clearly requires a vast amount of information. Yet our universe is strikingly simple in that the matter is nearly uniformly distributed on
the largest scales, with density fluctuations only at the $10^{-5}$ level.
Because of gravity, clumpier distributions are exponentially more likely.
Even on small scales, where gravitational effects are less important, there are
many regions that are very far from thermal equilibrium and heat death, \eg, the volume containing your brain.
This has disturbed many authors, including Boltzmann, who pointed out that a possible 
explanation of both his past perceptions and the arrow of time was that he was a 
disembodied brain that had temporarily assembled as a thermal fluctuation, with all his memories in place.
Yet this was clearly not the case, as it made a falsifiable prediction for his future perceptions: 
rapid disintegration towards heat death.

The most promising process to emerge that may help resolve this paradox is eternal cosmological inflation \cite{Guth81,Vilenkin83,Starobinsky84,LindeBook,Guth07}.
It makes the generic post-inflationary region rather homogeneous just as observed, thereby explaining why the entropy we observe is so low.\footnote{Because of issues related
to what, if anything, preceded inflation, it is still hotly debated whether eternal inflation completely solves the problem or is merely an ingredient of a yet 
to be discovered solution \cite{Penrose89,HollandsWald02,KofmanLindeMukhanov02,CarrollChen05}.
}
Although this has been less emphasized in the literature \cite{nihilo}, 
{\it inflation also explains why the entropy we observe is so high}: 
even starting with a state with so low complexity that it can be defined on a single page (the Bunch-Davies vacuum of quantum field theory), 
decoherence produces vast numbers of for all practical purposes parallel universes that generically exhibit complexity comparable to ours
\cite{nihilo,ZehBook,PolarskiStarobinsky96,KieferPolarski98}.

Boltzmann's paradox becomes very clear in the context of the MUH.
Consider, for example, the mathematical structure of classical relativistic field theory, \ie, 
a number of classical fields in spatially and temporally infinite Minkowski space,
with no initial conditions or other boundary conditions specified. 
This mathematical structure is the set of {\it all} solutions to the field equations. These different solutions are not in any way connected, 
so for all practical purposes, each constitutes its own parallel universe with its own ``initial conditions''.
Even if this mathematical structure contains SAS's (``observers''), it is ruled out as a candidate for describing our world,
independently of the fact that it lacks quantum mechanical and general relativity effects.
The reason is that generic solutions are a high-entropy mess, so that 
generic observers in this structure are of the above-mentioned disembodied-brain type,
\ie, totally unstable. 
The general unease that Boltzmann and others felt about this issue thus sharpens into a clear-cut conclusion in the MUH 
context:  this particular mathematical structure is ruled out as a candidate for the one we inhabit.

The exact same argument can be made for the mathematical structure corresponding to classical general relativity without a 
matter component causing inflation.
Generic states emerging from a Big-Bang like singularity are a high-entropy mess, so that 
almost all observers in this structure are again of the disembodied-brain type. This is most readily seen by evolving a generic state forward until a 
gigantic black hole is formed; the time-reversed solution is then ``a messy  big bang''.
In contrast, a mathematical structure embodying general relativity and inflation may be consistent with what we observe, 
since generic regions of post-inflationary space are now quite uniform.\footnote{Rigorously defining ``generic'' in the 
context of eternal inflation with infinite volumes to compare is still an open problem; see, \eg, 
\cite{inflation,Easther06,Bousso06,Vilenkin06,Aguirre06} for recent reviews.}

\subsubsection{How complex is our world?}
\label{ComplexWorldSec}

\begin{figure}
\centerline{{\vbox{\epsfxsize=8.5cm\epsfbox{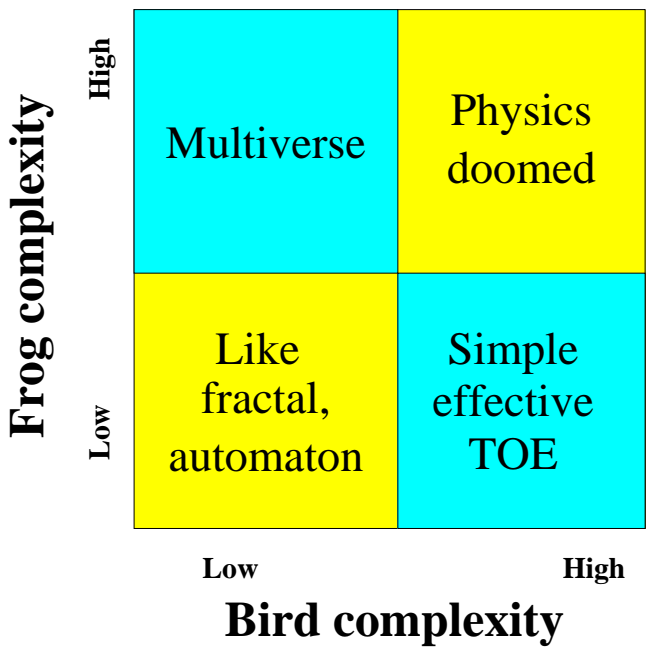}}}}
\smallskip
\caption{The four qualitatively different possibilities regarding the cosmic complexity
of the world. If a (bird's view) theory of everything can be described by fewer bits than 
our (frog's view) subjectively perceived universe, then we must live in a multiverse.
}
\label{ComplexityFig}
\end{figure}

\Fig{ComplexityFig} illustrates the four qualitatively different answers to this question,
depending on the complexity of a complete description of the frog's (inside) view and
the bird's (outside) view.

As mentioned above, the frog complexity is, taken at face value, huge.
However, there is also the logical possibility that this is a mere illusion, just as the Mandelbrot set 
and various patterns generated by cellular automata \cite{Wolfram02} can appear
complex to the eye even though they have complete mathematical descriptions that are very simple.
I personally view it as unlikely that all the observed star positions and other numbers that characterize our universe
can be reduced to almost nothing by a simple data compression algorithm. Indeed, cosmological 
inflation explicitly predicts that the seed fluctuations are distributed like Gaussian random variables, 
for which it is well-known that such dramatic compression is impossible.\footnote{The quantum fluctuations generated by inflation have a multivariate Gaussian distribution, 
which means that transforming
to the eigenbasis of the covariance matrix and scaling by the square roots of its eigenvalues converts the data 
into independent Gaussian variables with zero mean and unit standard deviation.
This implies that after transforming these Gaussian variables to ones with a uniform distribution, no further data compression is possible, 
so that the total complexity of the initial conditions is huge and in the same ballpark as one would naively guess.
}

The MUH does not specify whether the complexity of the mathematical structure in the bird perspective is low or high, 
so let us consider both possibilities.
If it is extremely high, our TOE quest for this structure is clearly doomed. In particular, if describing the structure 
requires more bits than describing our observable universe, it is of course impossible to store the information about 
the structure in our universe.
If it is high but our frog's view complexity is low, then we have no practical need for 
finding this true mathematical structure, since the frog description provides a simple effective TOE
for all practical purposes. 

A widely held hope among theoretical physicists is that the bird complexity is in fact very low, so that the
TOE is simple and arguably beautiful. There is arguably no evidence yet against this simplicity hypothesis \cite{nihilo}.
If this hypothesis is true and the bird complexity is much lower than the frog 
complexity, then {\it it implies that the mathematical structure describes some form of multiverse}, 
with the extra frog complexity entering in describing which parallel universe we are in.
For example, if it is established that a complete (frog's view) description of the current state of our observable universe 
requires $10^{100}$ bits of information, then either the (bird's view) mathematical structure requires 
$\ge 10^{100}$ bits to describe, or we live in a multiverse. 
As is discussed in more detail in, \eg, 
\cite{nihilo,Schmidhuber00,StandishBook},
{\it an entire ensemble is often much simpler than one of its members}.
For instance, the algorithmic information content \cite{Chaitin90,LiVitanyiBook}
in a number
is roughly speaking defined as the length (in bits)
of the shortest computer program
which will produce that number as output, so the information
content in a generic integer $n$ is of order $\log_2 n$.
Nonetheless, the set of all integers $1, 2, 3, ...$
can be generated by quite a
trivial computer program,
so the algorithmic
complexity of the whole set is smaller than that of a generic member.
Similarly, the set of all perfect fluid solutions to the
Einstein field equations has a smaller algorithmic complexity than
a generic particular solution, since the former is characterized simply by
giving the Einstein field equations and the latter requires the specification of
vast amounts of initial data on some spacelike hypersurface.
Loosely speaking, the apparent information content rises when
we restrict our attention to one particular element in an ensemble,
thus losing the symmetry and simplicity that was inherent in the totality
of all elements taken together.
The complexity of the whole ensemble is thus not only smaller than the sum of that of its parts, 
but it is even smaller than that of a generic one of its parts.

\subsection{Physical constants}

Suppose the mathematical structure has a finite complexity, \ie, can be defined with a finite number of bits.
This follows from the Computable Universe Hypothesis proposed in \Sec{GoedelSec}, but also follows for any other ``elegant universe'' case
where the TOE is simple enough to be completely described on a finite number of pages.

This supposition has strong implications for physical constants. 
In traditional quantum field theory, the Lagrangian contains dimensionless parameters that 
can in principle take any real value. Since even a single generic real number requires an infinite 
number of bits
to specify, no computable finite-complexity mathematical structure can have such a Lagrangian.
Instead, fundamental parameters must belong to the countable set of numbers that are specifiable with a finite amount of information,
which includes integers, rational numbers and algebraic numbers.
There are therefore only two possible origins for random-looking parameters in the standard model Lagrangian
like $1/137.0360565$: either they are computable from a finite amount of information,
or the mathematical structure corresponds to a multiverse where the parameter takes {\it each} real number 
in some finitely specifiable set in some parallel universe.
The apparently infinite amount of information contained in the parameter then merely reflects there being an uncountable number of 
parallel universes, with all this information required to specify which one we are in.
The most commonly discussed situation in the string landscape context is a hybrid between these two: the
parameters correspond to a discrete (finite or countably infinite) set of solutions to some equation, each set defining 
a stable or metastable vacuum which is for all practical purposes a parallel universe.

\clearpage
\begin{figure}[tbp]
\vskip-1.5cm
\centerline{{\vbox{\hglue-0.5cm\epsfxsize=19.4cm\epsfbox{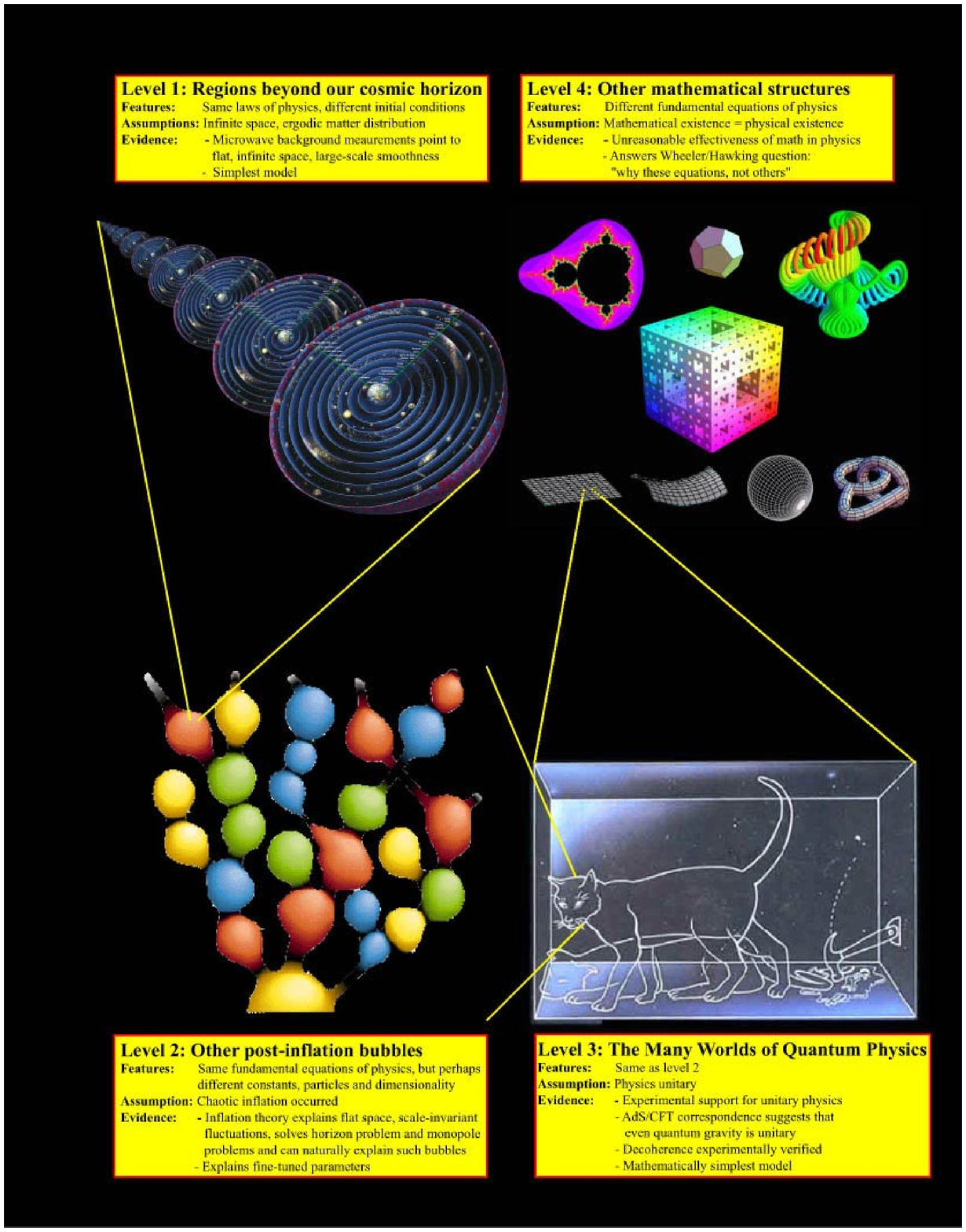}}}}
\label{ZoomFig}
\end{figure}
\setcounter{figure}{4}
\clearpage

\section{Implications for parallel universes}
\label{PUsec}

Parallel universes are now all the rage, cropping up in books, movies and 
even jokes: ``You passed your exam in many parallel universes --- but not this one."
However, they are as controversial as they are popular, and it is important to ask whether they are 
within the purview of science, or merely silly speculation. 
They are also a source of confusion, since many forget to distinguish between 
different types of parallel universes that have been proposed, whereas 
\cite{multiverse,multiverse4wheeler} argued that the various proposals form
a natural four-level hierarchy of multiverses (\fig{ZoomFig}) allowing progressively greater diversity: .
\begin{itemize}
\item {\bf Level I:} A generic prediction of cosmological inflation is an infinite ``ergodic'' space,
which contains Hubble volumes\footnote{Our {\it Hubble volume} is defined as the spherical 
region from which light has had time to 
reach us during the 14 billion years since our big bang. This sphere is also referred to as 
our {\it horizon volume} or simply our universe.}
realizing all initial conditions \cite{GarrigaVilenkin01,multiverse} --- including
an identical copy of you about $10^{10^{29}}$m away \cite{multiverse}.
\item {\bf Level II:} Given the mathematical structure corresponding to 
the {\it fundamental} laws of physics 
that physicists one day hope to capture with equations on a T-shirt, 
different regions of space can exhibit different
{\it effective} laws of physics 
(physical constants, dimensionality, particle content, \etc) corresponding
to different local minima in a landscape of possibilities. 
\item {\bf Level III:} In unitary quantum mechanics \cite{Everett57,EverettBook}, other branches of the wave function 
add nothing qualitatively new, which is ironic given
that this level has historically been the most controversial.
\item {\bf Level IV:} Other mathematical structures give different fundamental equations of physics for that T-shirt.
\end{itemize}
The key question is therefore not whether there is a multiverse (since Level I is the rather uncontroversial 
cosmological standard model), but rather how many levels it has.

\subsection{Physics or philosophy?}

The issue of evidence and whether this is science or philosophy has been discussed at length in the recent literature --- see, 
\eg, \cite{multiverse,multiverse4wheeler,DeutschBook,Linde02,Ellis04,Stoeger04,ReesHabitat,Holder04,Weinberg05,PageNonsense,Carroll06,Aguirre06,DaviesUniversesGalore,KakuBook,VilenkinBook}.
The key point to remember is that
{\it parallel universes are not a theory, but a prediction of certain theories}.
For a theory to be falsifiable, we need not be able to observe and test all its predictions, merely at least one of them.
Consider the following analogy:
\begin{center}
\begin{tabular}{|l|l|}
\hline
{\it General Relativity}      	&{\it Black hole interiors}\\
\hline
{\it Inflation}               	&{\it Level I parallel universes}\\
\hline
{\it Inflation$+$landscape}	&{\it Level II parallel universes}\\
\hline
{\it Unitary quantum mechanics}&{\it Level III parallel universes}\\
\hline
{\it MUH}			&{\it Level IV parallel universes}\\
\hline
\end{tabular}
\end{center}
Because Einstein's theory of General Relativity has successfully predicted many things that we {\it can} observe, we also take seriously its predictions for things we
cannot observe, \eg, that space continues inside black hole event horizons and that (contrary to early misconceptions) nothing funny happens
right at the horizon.
Likewise, successful predictions of the theories of cosmological inflation and 
unitary
quantum mechanics have made some scientists take more seriously their other predictions, 
including various types of parallel universes.

\subsection{Levels I, II and III}
  
Level I, II and III parallel universes are all part of the same 
mathematical structure, but from the frog perspective, they are for all practical purposes 
causally disconnected. Level II, is currently a very active research area.
The possibility of a string theory ``landscape'' 
\cite{Bousso00,Feng00,KKLT03,Susskind03,AshikDouglas04},
where the above-mentioned potential has perhaps $10^{500}$ different minima, may offer a specific realization of
the Level II multiverse which would in turn have four sub-levels of increasing diversity:
\begin{itemize}
\item {\bf IId:} different ways in which space can be compactified, which can allow both different effective
dimensionality and different symmetries/elementary articles (corresponding to different 
topology of the curled up extra dimensions).
\item {\bf IIc:} different ``fluxes'' (generalized magnetic fields) that stabilize the extra dimensions (this sublevel is where the largest 
number of choices enter, perhaps $10^{500}$).
\item {\bf IIb:} once these two choices have been made, there may be 
multiple
minima in the effective supergravity potential.
\item {\bf IIa:} the same minimum and effective laws of physics can be realized in many different post-inflationary bubbles, each 
constituting a Level I multiverse.
\end{itemize}
Level I is relevant to our discussion above in \Sec{ICsubsec}, since the density fluctuations from cosmological inflation 
realize all initial conditions with a well-defined statistical distribution. 
Level III is relevant to the details of this generation process, since the quantum superposition of field 
configurations decoheres into what is for all practical purposes an ensemble of classical universes with different 
density fluctuation patterns.
Just as in all cases of spontaneous symmetry breaking, the symmetry is never broken in the bird's view, merely in the 
frog's view: a translationally and rotationally quantum state (wave functional) such as 
the Bunch-Davies vacuum can decohere into an incoherent superposition of states that lack any symmetry.
I suspect that we would be much less confused by Everett's Level III ideas if we were computers and 
familiar with the subjective aspects of routinely making and deleting copies of ourselves.

In summary, many mathematical structures contain {\it de facto} parallel universes at 
levels I through III, so the possibility of a multiverse is a 
direct and obvious implication of the MUH. We will therefore not dwell further on 
these levels, and devote the remainder of this section to Level IV.

\subsection{Motivation for Level IV}

If the TOE at the top of \fig{TheoryTreeFig} exists and is one day discovered, 
then an embarrassing question remains, as emphasized by John Archibald
Wheeler: {\it  Why these particular equations, not others?}
Could there really be a fundamental, unexplained ontological asymmetry built into the
very heart of reality, splitting mathematical structures into two classes,
those with and without physical existence?
After all, a mathematical structure is not ``created'' and doesn't exist ``somewhere''. 
It just exists.\footnote{This paper uses the term ``mathematics" as defined by 
the formalist school of mathematicians, according to which it is by construction human-independent
and hence has the Platonic property of being ``out there" out be discovered rather than invented.
It is noteworthy that there is a long history of contention among matematicians and philosophers
as to which is the most appropriate definition of mathematics for mathematicians to use
(see \cite{Feferman,Hersh} for recent updates). 
}

As a way out of this philosophical conundrum, I have suggested \cite{toe})
that complete mathematical democracy holds: that mathematical existence and 
physical existence are equivalent, so that {\it all} mathematical structures have the same ontological status.
This can be 
viewed as a form of radical Platonism,
asserting that the mathematical structures
in Plato's {\it realm of ideas}, the {\it Mindscape} of Rucker \cite{Rucker82}, exist 
``out there'' in a physical sense \cite{DaviesGod},
casting the so-called modal realism theory of  
David Lewis \cite{Lewis86} in mathematical terms
akin to what Barrow \cite{BarrowTOE,BarrowPi} refers to as ``$\pi$ in the sky''.
If this theory is correct, then since it has no free parameters,
all properties of all parallel universes
(including the subjective perceptions of SAS's in them) could in principle be derived by
an infinitely intelligent mathematician.

{\it In the context of the MUH, the existence of the Level IV multiverse is not optional.}
As was discussed in detail in \Sec{DescriptionSec}, the MUH says that 
a mathematical structure {\it is} our external physical reality, rather than being merely a
{\it description} thereof.
This equivalence between physical and mathematical existence means that 
if a mathematical structure contains a SAS, it will 
perceive itself as existing in a physically real world, just as we do 
(albeit generically a world with different properties from ours).
Stephen Hawking famously asked 
``what is it that breathes fire into the equations and
makes a universe for them to describe?'' \cite{Hawking93}.
In the context of the MUH, there is thus no breathing required, 
since the point is not that a mathematical structure {\it describes} a universe, but that it {\it is} a universe.

\subsection{The ``anything goes'' critique}

The Level IV multiverse idea described in \cite{toe} has been extensively discussed in the recent literature 
(see, \eg, 
\cite{Ellis99,Schmidhuber00,CSchmidhuber00,Hogan00,Benioff00,Ellis02,Linde02,BostromBook,Benioff02,Benioff03,Circovic03,Vaas04,conditionalization,Benioff05,McCabe05b,mmm,Vorhees05,Ellis04,Stoeger04,DaviesUniversesGalore,Holder04,StandishBook,Ellis06,Stoeger06,Hedrich06}),
and has been criticized on several grounds which we will now discuss.

One common criticism asserts that it implies that all imaginable universes exist, and therefore makes no predictions at all.
Variations of this have also been used as a theological argument, e.g., 
``All imaginable universes exist, so there's at least one with an omnipotent God who, since he's omnipotent, can 
insert himself into all other structures as well. Therefore God exists in our universe.''
Milder claims are also common, \eg,
``at least some of these universes will feature miraculous events - water turning into wine, {\etc}'' \cite{DaviesUniversesGalore}.

The MUH and the Level IV multiverse idea does certainly {\it not} imply that all imaginable universes exist.
We humans can imagine many things that are mathematically undefined and hence do not correspond to mathematical structures.
Mathematicians publish papers with existence proofs and demonstrating the mathematical consistency of various mathematical structures
precisely because this is difficult and not possible in all cases.

A closely related misconception relates to the issue of ``baggage'' described in \Sec{BaggageSec}.
If a mathematical structure is objects with 
relations between them, then is not {\it any} theory a mathematical 
structure?
The answer is no: unless it can be defined in the form specified in Appendix A, 
with tableaus or algorithms explicitly defining all relations, it is not a mathematical structure.
Consider the proposition ``God created Adam and Eve'', where there are entities ``God'', ``Adam'' and ``Eve'' linked by a relation 
denoted ``created''.
When we humans read this proposition, words like ``God'' come with baggage, 
\ie, additional properties and connotations not explicitly spelled out. 
To distill this proposition into a mathematical structure, we need to 
clear the names of the set elements and relations 
of any intrinsic meaning, so it is equivalent to saying simply
that there is a set $S$ with three elements $s_i$ and a relation  
such that $R(s_i,s_j)$ holds if and only if $i=1$ and $j\ge 2$.
To us humans, this latter description has a very different feel to it,  
since $s_1$ has no connotations of omnipotence, ``ability to create universes'', {\etc}
An English language statement involving relations is thus not a mathematical structure unless we either 
sacrifice all connotations or add further entities and relations that somehow encode them.

Regarding the above-mentioned water-into-wine issue \cite{DaviesUniversesGalore}, the known Poincar\'e-invariant laws of physics
allow ``miraculous'' events to occur via thermal fluctuations or quantum tunneling, but with exponentially small probabilities that for
all practical purposes preclude us from observing them. It is far from obvious that there exist other mathematical structures where 
such ``miracles'' are common while regularities remain sufficient to support the evolution of sentient observers, especially considering the
fact that the cosmological measure problem \cite{inflation,Easther06,Bousso06,Vilenkin06,Aguirre06} (of how to compute what is common and rare
in an inflationary landscape) remains unsolved; we return to this issue below.

\subsection{Other critique}

A separate critique of the Level IV idea has recently been leveled by Page \cite{PageNonsense}, who argues that it is 
{\it ``logical nonsense to talk of Level IV in the sense of the co-existence of all mathematical structures''}.
This criticism appears to be rooted in a misconception of the notion of mathematical structure, as \cite{PageNonsense} goes on to say
{\it ``Different mathematical structures can be contradictory, and contradictory ones cannot co-exist. For example, one structure could 
assert that spacetime exists somewhere and another that it does not exist at all.
However, these two structures cannot both describe reality.''}
Being baggage free, mathematical structures of course ``assert'' nothing. Different structures are, by definition, different, so one may 
contain Minkowski space while another does not, corresponding to two very different Level IV parallel universes.
Page also postulates that ``There must be one unique mathematical structure that describes ultimate reality.'' \cite{PageNonsense}.
As we will see in the following section, this is less inconsistent with Level IV than it may sound, since many mathematical structures 
decompose into unrelated substructures, and separate ones can be unified.

Interesting potential problems related to G\"odel incompleteness and measure have been raised by 
Vilenkin \cite{VilenkinBook} and others, and we will devote the entire \Sec{GoedelSec} below to this fascinating topic. 
Here we merely note that since the measure problem has not even been solved at 
Level II yet \cite{inflation,Easther06,Bousso06,Vilenkin06,Aguirre06}, it can hardly be considered a proven show-stopper specifically for Level IV at
this point.

\subsection{Disconnected realities}
  
The definition of a mathematical structure $S$ as abstract entities with relations between them
(see Appendix A for mathematical details) leads to a natural decomposition.
Let us define a mathematical structure $S$ as {\it reducible} if its entities can be split into two sets 
$T$ and $U$ with no relations relating the two.
For example, there must be no relations like $R(t_i)=u_j$ or $R(t_i,u_j)=t_k$ for any elements 
$t_i,t_k\in T$ and $u_j\in U$. Then $T$ and $U$ individually satisfy the definition of a mathematical structure.
Since the relations are the {\it only} properties that the set elements have, 
and the elements of the structures $U$ and $T$ are by definition unrelated, these mathematical structures $U$ and $T$
are parallel Level IV universes as far as an observer in either of them is concerned.

As a trivial example, consider the mathematical structure corresponding to the set of $n$ elements with no relations.
This decomposes into $n$ structures that are the set of one element.
In contrast, most mathematical structures discussed in Appendix A (\eg, Boolean algebra, the integers, and Euclidean space) are irreducible, \ie, 
cannot be decomposed. 
The minimal mathematical structure containing the Level I, Level II and Level III multiverses is also irreducible.
For example, there are mathematically defined relations between points in causally separated Hubble volumes (generated by relations between topologically 
neighboring points) even though they are inaccessible to observers.
Similarly, there are mathematically defined relations such as the inner product between pairs of states in a quantum Hilbert space even though decoherence renders them 
for all practical purposes parallel Level III universes.

Given two mathematical structures, one can always define a third one that unifies them: the reducible structure consisting of the union of all 
elements from both, with their corresponding relations. 
Thus any finite number of Level IV parallel universes can be subsumed within one single mathematical structure, along the lines 
of Page's above-mentioned preference \cite{PageNonsense}. We will discuss subtleties related to infinite numbers of structures below in 
\Sec{GoedelSec}.
The inside view of a given irreducible mathematical structure is thus the same regardless of whether it is defined alone or as part of 
a larger reducible mathematical structure. The only instance when this distinction might matter is if some sort of measure over the Level IV multiverse
is involved, an issue to which we return in \Sec{GoedelSec}.

As pointed out by Linde \cite{Linde02}, one can implement even as a single Level III multiverse 
the subset of the Level IV multiverse corresponding to 
mathematical structures that are similar to ours in being described by Lagragrangians: 
one simply defines a single Lagrangian which is the sum of Lagrangians in disjoint variables.

\section{Implications for the simulation argument}
\label{SimSec}

Long a staple of science fiction, the idea that our external reality is some form 
of computer simulation has gained prominence with recent blockbuster movies like
{\it The Matrix}. 

\subsection{Are we simulated?} 

Many scientists have argued that simulated minds are both possible and imminent
(\eg, \cite{DrexlerSimo,Bostrom98,Kurzweil99,Moravec99}), 
and, \eg, Tipler \cite{Tipler}, Bostrom \cite{BostromSimo} and Schmidhuber \cite{Schmidhuber97,Schmidhuber00} have
gone as far as discussing the probability that we are simulated
(this further complicates the above-mentioned measure problem). 
On the other hand, McCabe has presented an argument for why 
a self-aware being can never be simulated by any form of digital computer \cite{McCabe05}.\footnote{It is noteworthy that computations in
our brains are discrete to the extent that neuron firings are, and that 
even binary-like representations appear to be employed in some contexts. 
For example, so-called grid cells 
roughly speaking record our position modulo a fixed reference length \cite{Hafting05},
so a collection of grid cell layers whose reference lengths vary by powers of two would effectively store 
the binary digits of our position coordinates.}
This raises a question emphasized by Penrose \cite{PenroseBook89,Penrose97}: is it possible to build a human impersonator that passes the Turing test,
yet is in no sense sentient? 

Lloyd has advanced the intermediate possibility that we live in an analog simulation performed by a quantum computer, 
albeit not a computer designed by anybody --- rather, because the structure of quantum field theory is mathematically equivalent 
to that of a spatially distributed quantum computer \cite{Lloyd97}.
In a similar spirit, Schmidhuber \cite{Schmidhuber00}, Wolfram \cite{Wolfram02} and others have explored the idea that the laws of physics correspond to a classical computation.
Below we will explore these issues in the context of the MUH.

\subsection{The time misconception}
\label{TimeMisconceptionSec}

Suppose that our universe is indeed some form of computation.
A common misconception in
the universe simulation literature is that our physical notion of a one-dimensional time must then necessarily be equated 
with the step-by-step one-dimensional flow of the computation.
I will argue below that if the MUH is correct, then computations do not need to {\it evolve} the universe, but merely 
{\it describe} it (defining all its relations).

The temptation to equate time steps with computational steps is understandable, given that both form a one-dimensional 
sequence where (at least for the non-quantum case) the next step is determined by the current state.
However, this temptation stems from an outdated classical description of physics:
there is generically no natural and well-defined global time variable in general relativity, and even less so in quantum 
gravity where time emerges as an approximate semiclassical property of certain ``clock'' subsystems (\eg, \cite{Gambini04}).
Indeed, linking frog perspective time with computer time is unwarranted even within the context of classical physics. 
The rate of time flow perceived by an observer in the simulated universe is completely independent of the rate at which a computer runs the 
simulation, a point emphasized in, \eg, \cite{Egan95,toe,Schmidhuber00}.
Moreover, as emphasized by Einstein, it is arguably more natural to view our universe not from the frog perspective as a 3-dimensional space
where things happen, but from the bird perspective as a 4-dimensional spacetime that merely is.
There should therefore be no need for the computer to compute anything
at all --- it could simply store all the 4-dimensional data, \ie, encode all properties of the mathematical structure that is our universe.
Individual time slices could then be read out sequentially if desired, and 
the ``simulated" world should still feel as real to its inhabitants as in the case where only 3-dimensional data is stored and evolved
\cite{toe}. In conclusion, the role of the simulating computer is not to compute the history of our universe, but to specify it.

How specify it? 
The way in which the data are stored (the type of computer, data format, \etc) should be irrelevant, so 
the extent to which the inhabitants of the simulated universe perceive themselves as real 
should be invariant under data compression. 
The physical laws that we have discovered provide great
means of data compression, since they make it sufficient to store the initial 
data at some time together with the equations and an integration routine.
As emphasized in \cite{nihilo} and \Sec{ComplexitySec} above, the initial data might be extremely simple: quantum field theory states 
such as the Hawking-Hartle wave function or the inflationary Bunch-Davies vacuum have very low algorithmic complexity 
(since they can be defined in quite brief physics papers), yet simulating their time evolution would simulate not merely 
one universe like ours, but a vast decohering ensemble corresponding to the above-mentioned Level III multiverse.
It is therefore plausible that our universe could be simulated by quite a short computer 
program \cite{nihilo,toe,Schmidhuber00}.

\subsection{A different sort of computation}

The above example referred to our particular mathematical structure.
More generally, as we have discussed, 
a complete description of 
an arbitrary mathematical structure $S$ is by definition a specification of the relations between the elements of $S$.
For finite structures, this specification can take the form of finite tableaus of numbers akin to generalized multiplication tables. 
More generally, they are functions whose arguments take infinitely many values and hence cannot be tabulated. 
For these relations to be well-defined, all these functions must be {\it computable}, \ie, there must exist a computer program that will 
take (some bit string encoding of) the arguments as input and, in a finite number of computational steps, output the
(bit string encoding of) the function value.
Each relation of the mathematical structure is thus defined by a computation.
In other words, if our world is a well-defined mathematical structure in this sense, then
it is indeed inexorably linked to computations, albeit computations of a different sort than those
usually associated with the simulation hypothesis:
these computations do not {\it evolve} the universe, but {\it define} it by evaluating its relations.

\subsection{The transcendent structure}

\begin{figure}[pbt]
\centerline{{\vbox{\epsfxsize=8.7cm\epsfbox{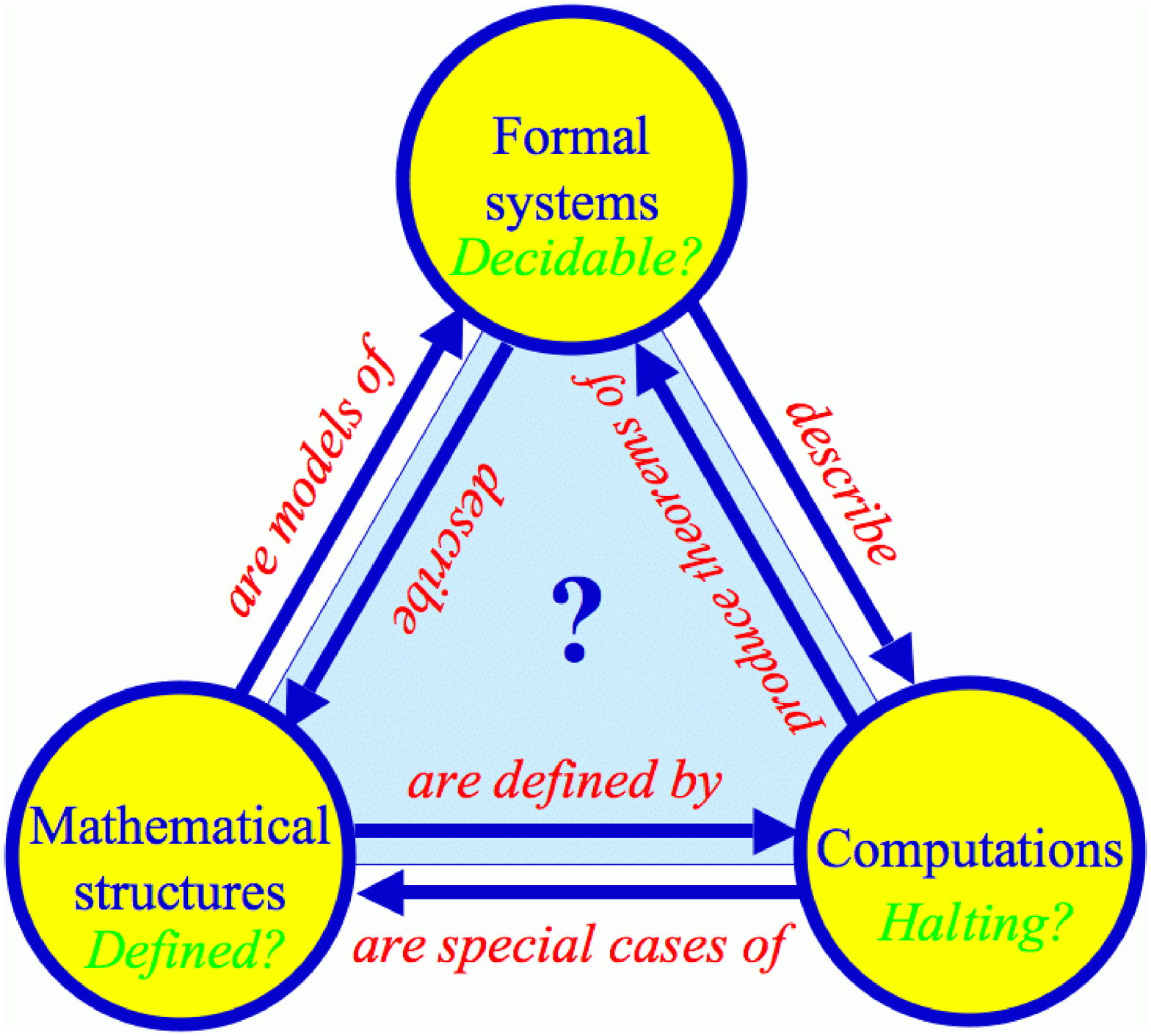}}}}
\smallskip
\caption{The arrows indicate the close relations between mathematical structures, 
formal systems, and computations.
The question mark suggests that these are all aspects of the same
transcendent structure (the Level IV multiverse), and that we still have not
fully understood its nature.
}
\label{StructureFig}
\end{figure}

Above we discussed how mathematical structures and computations are closely related, 
in that the former are defined by the latter.
On the other hand, computations are merely special cases of mathematical structures. 
For example, the information content (memory state) of a digital computer is a string of bits, say 
``$1001011100111001...$'' of great but finite length, equivalent to some large 
but finite integer $n$ written in binary. The information processing of a computer is
a deterministic rule for changing each memory state into another (applied over and over again),
so mathematically, it is simply a function $f$ mapping the integers onto themselves
that gets iterated: $n\mapsto f(n)\mapsto f(f(n))\mapsto...$.
In other words, even the most sophisticated computer simulation is
merely a special case of a mathematical structure, hence included in the Level IV multiverse.

\fig{StructureFig} illustrates how computations and mathematical structures are related not only to each other, but also
to formal systems, which formally describe them. 
If a formal system describes a mathematical structure, the latter is said to be a model of the former \cite{Hodges}.
Moreover, computations can generate theorems of formal systems (indeed, there are algorithms computing all theorems for 
the class of formal systems known as recursively enumerable). 
For mathematical structures, formal systems, and computations alike, there is a subset with 
an attractive property (being defined, decidable and halting, respectively), and 
in \Sec{GoedelSec} below, we will see that these three subsets are closely related.

I have drawn a question mark in the center of the triangle to suggest 
that the three mathematical structures, formal systems, and computations are simply different aspects of one
underlying transcendent structure whose nature we still do not fully understand. 
This structure (perhaps restricted to the defined/decidable/halting part as hypothesized in \Sec{GoedelSec} below)
exists ``out there'', and is both the totality of what has mathematical existence and 
the totality of what has physical existence, \ie, the Level IV multiverse.

\subsection{Does a simulation really need to be run?}

A deeper understanding of the relations between 
mathematical structures, formal systems, and computations would shed light on many of the issues raised in this paper.
One such issue is the above-mentioned measure problem, which is in essence the problem of how to 
deal with annoying infinities and predict conditional probabilities for what an observer should 
perceive given past observations.

For example, since every universe simulation corresponds to a mathematical structure,
and therefore already exists in the Level IV multiverse, does it in some meaningful sense exist ``more'' if it is 
in addition run on a computer? This question is further complicated by the fact that eternal inflation predicts 
an infinite space with infinitely many planets, civilizations, and computers, and that the Level IV multiverse includes an infinite number of possible simulations.
The above-mentioned fact that our universe (together with the entire Level III multiverse) may be simulatable by quite a short computer program
(\Sec{TimeMisconceptionSec}) calls into question whether it makes any ontological difference whether simulations are ``run'' or not.
If, as argued above, the computer need only describe and not compute the history, then the complete description would probably fit on a single memory stick, 
and no CPU power would be required. It would appear absurd that the existence of this memory stick would 
have any impact whatsoever on whether the multiverse it describes
exists ``for real''. Even if the existence of the memory stick mattered, some elements of this multiverse will contain an identical memory stick that would 
``recursively" support its own physical existence.
This would not involve any Catch-22 ``chicken-and-egg" problem
regarding whether the stick or the multiverse existed first, 
since the multiverse elements are 4-dimensional spacetimes, whereas
``creation" is of course only a meaningful notion {\it within} a spacetime.

We return to the measure problem in \Sec{CUHmeasureSec}, and issues related to the algorithmic complexity of 
mathematical structures, formal systems and computations.

\section{The Computable Universe Hypothesis}
\label{GoedelSec}

In this section, we will explore the consequences of augmenting the Mathematical Universe Hypothesis with
a second assumption:

\smallskip
\centerline{\framebox{\parbox{7cm}{
{\bf Computable Universe Hypothesis (CUH):} {\it The mathematical structure that is our external physical reality is
defined by computable functions.}
}}}
\smallskip

By this we mean that the relations (functions) that define the mathematical structure 
as in Appendix~\ref{StructureDefSec}
can all be 
implemented as computations that are guaranteed to halt after a finite number of steps.\footnote{This 
is not to be confused
with another common usage of the term ``computable function'', as functions that can be approximated 
to any desired accuracy in a finite number of steps (as opposed to computed exactly as in our definition). 
To distinguish these two usages, we can refer to our definition as an ``exactly computable function'' and to the alternative
usage as an ``approximately computable function''. Exact computability implies approximate computability, but the converse is not true.
For example, if a real-valued function $f$ with no arguments
(a constant) is approximately computable and we keep computing it to be consistent with zero
to ever greater accuracy, we still do not know whether $f<0$, $f=0$ or $f>0$. 
As described below, functions of a real variable are never exactly computable, and there are interesting complications even 
for functions of integers.
}
We will see that the CUH places extremely restrictive conditions on mathematical structures, 
which has attractive advantages as well as posing serious challenges.

\subsection{Relation to other hypotheses}

Various related ideas have been discussed in the literature --- for an excellent overview, see \cite{StandishBook}.
Schmidhuber has hypothesized that all halting programs (together with certain non-halting ones) 
correspond to physical realities \cite{Schmidhuber97,Schmidhuber00}, 
and related ideas have also been explored by, \eg , \cite{BarrowTOE}. % Page 205
Note that the CUH is a different hypothesis: it requires the {\it description} (the relations) rather than the {\it time evolution} to be computable.
In other words, in terms of the three vertices of \fig{StructureFig}, 
the universe is a mathematical structure according to the MUH, as opposed to a 
computation according to the simulation hypothesis. 

In addition, Barrow has suggested that 
only structures complex enough for G\"odel's incompleteness 
theorem to apply can contain self-aware observers \cite{BarrowPi}. Below we will see that the CUH in a sense postulates the exact opposite.

\subsection{Mathematical structures, formal systems, and computations}

Before exploring the implications of the CUH, let us discuss the motivation for it.

I have long wondered whether G\"odel's incompleteness theorem in some sense torpedos the MUH, and 
concerns along these lines have also been raised by 
Vilenkin \cite{VilenkinBook} and others. % Page?
Indeed, this worry is somewhat broader, extending to all three vertices of \fig{StructureFig} above.
This figure illustrates that mathematical structures, formal systems, and computations are closely related, suggesting 
that they are all aspects of the same
transcendent structure (the Level IV multiverse), whose nature we have still not fully understood.
The figure also illustrates that there are potential problems at all three vertices of the triangle:
mathematical structures may have relations that are undefined, 
formal systems may contain undecidable statements,  
and computations may not halt after a finite number of steps. 
Let us now revisit the relation between these three issues in more detail than we did above in \Sec{SimSec}.

The relations between the three vertices with their corresponding complications are illustrated by six arrows.
Since different arrows are studied by different specialists in a range of fields from 
mathematical logic 
to computer science,
the study of the triangle as a whole is somewhat interdisciplinary. % Let us therefore briefly review some key points.
The connection between formal systems and mathematical structures is the subject of the branch of mathematics known as 
model theory (see, \eg, \cite{Hodges}).
The use of computations to automatically prove theorems of formal systems is actively studied by computer scientists, and arbitrary
computations are in turn special cases of mathematical structures that are described by formal systems. 
Finally, as we have discussed, mathematical structures can be explicitly defined by computations that implement their relations.
Further discussion of the relation between mathematical structures and computations can be found in, \eg, \cite{toe,Standish00,multiverse4wheeler}.

Note that the correspondence between the three vertices is not one-to-one.
There are often many mathematical structures that are described by (``are models of'') a given formal system; for example, 
all specific groups are models of the formal system containing only the group axioms.
Conversely, several formal systems may be equivalent descriptions of the same mathematical structure --- {\eg} \cite{toe} gives examples of this.
Only a subset of all mathematical structures can be interpreted as computations, 
simple examples include 
certain integer mappings (which can be interpreted as defining 
a Universal Turing Machine or other digital computer \cite{DavisComputability}). 
Our world can also be thought of as a sort of computation \cite{Lloyd97}, since it allows computers to be built within it.
It appears not to support universal computation, however, as it lacks both unlimited storage capacity and 
time for an unlimited number of computational steps ({\cf} \cite{Tipler}).

Finally, many different sets of computations can define the same mathematical structure
as detailed in Appendix~\ref{StructureDefSec}, by implementing equivalent choices of generating relations and/or by corresponding to 
different labelings of the set elements.

\subsection{Undefined, undecidable and uncomputable}

Bearing these correspondences in mind, let us now discuss the G\"odel-inspired worry mentioned above:
that the MUH makes no sense because our universe would be somehow inconsistent or undefined.
If one accepts David Hilbert's dictum  that
{\it ``mathematical existence is merely freedom from contradiction''} \cite{Hilbert}, then 
an inconsistent structure would not exist mathematically, let alone physically as in the MUH.

Our standard model of physics includes everyday mathematical structures such as the integers (defined by the Peano axioms) 
and real numbers. Yet G\"odel's second incompleteness theorem \cite{Goedel} implies that we can never be  
$100\%$ sure that this everyday mathematics is consistent: it leaves open the possibility that 
a finite length proof exists within number theory itself demonstrating 
that $0=1$.\footnote{Note that it is not the inability of Peano Arithmetic (PA) to prove its own consistency that is the crux:
proving the consistency of PA using PA would still leave open the possibility that PA was inconsistent,
since an inconsistent theory implies all statements.
David Hilbert was of course well aware of this when he posed as his famous ``second problem" to prove the consistency of PA,
hoping instead that such a proof may be possible starting with a simpler (preferably finitistic) formal system 
whose consistency skeptics might be more likely to accept on faith. 
G{\"o}del's work torpedoed this hope, since one can only prove the consistency of PA using even 
stronger (and arguably still more suspect) axiom systems, involving, \eg, transfinite induction
or Zermelo-Frankel Set Theory, and consistency proofs for these systems require still stronger axiom systems.
The interpretation of these facts remains controversial --- see \cite{Simpson88,Dawson06} for two different viewpoints.
}
Using this result, every other well-defined statement in the formal system could in turn be proven to be true 
and mathematics as we know it would collapse like a house of cards \cite{Goedel}.

G\"odel's theorems refer to the top vertex in \fig{StructureFig}, \ie, formal systems,
and we must understand how this relates to lower left (the mathematical structures of the MUH) and the lower right (the computations 
that define the structures). 
Consider the example of number theory with predicate calculus, where  
G\"odel's first incompleteness theorem shows that there are undecidable statements which can neither be proved nor refuted.
Following the notation of Appendix A, the corresponding mathematical structure involves two sets: 
\begin{enumerate}
\item The two-element (Boolean) set $S_1\equiv\{0,1\}$ which we can interpret as $\{False,True\}$ and 
\item the countably infinite set $S_2\equiv\Integers$ which we can interpret as the integers.
\end{enumerate}
Standard logical operations such as $\sim$ (``not'') and $\&$ (``and'') are 
Boolean-valued functions of Boolean variables
(for example, $\&$ maps $S_1\times S_1\mapsto S_1$).
Operations such as $+$ and $\cdot$ map integers into integers ($+$ maps $S_2\times S_2\mapsto S_2$).
So-called predicates are functions mapping into Booleans that can mix the two sets by also taking integers as arguments.
A simple example is $<$ (``less than''), mapping $S_2\times S_2\mapsto S_1$.
In number theory, the quantifiers $\forall$ and $\exists$ allow the construction of more interesting functions,
such as the predicate $P$ mapping $S_2\mapsto S_1$ which is true if its argument is prime:
\beq{PrimeEq}
P(n) \equiv (\forall a)(\forall b)[[a>1]\och[b>1] \imp a\cdot b \neq n].
\eeq
This function $P(n)$ is undeniably well-defined in the sense that it is computable:
there exists a simple computer program that takes an arbitrary integer $n$ (represented as a bit string) 
as input, computes the single bit representing $P(n)$ in a finite number of steps, and then halts.
However, for the predicate $T(n)$ defined by 
\beq{TwinPrimeEq}
T(n) \equiv (\exists a)[(a>n)\och P(a)\och P(a+2)],
\eeq
no such halting algorithm for its computation has yet been discovered, since
it corresponds to the statement that there are twin primes larger than $n$.
One can write a simple computer program that loops and 
tests $a=n+1, n+2, ...$ and halts returning 1 if it finds $a$ and $a+2$ to be prime, 
and this program is guaranteed to correctly evaluate $T(n)$
as long as $n<2003663613\times 2^{195000}-1$, corresponding to the record twin prime
as of writing. However, since the existence of infinitely many twin primes is a famous open question, 
the program might not halt for all $n$, and no halting algorithm for evaluating $T(n)$ is currently known.
Equivalently, there is no known halting algorithm for evaluating the Boolean constant (function with no arguments)
\beq{TwinPrimeEq2}
T \equiv (\forall n)(\exists a)[(a>n)\och P(a)\och P(a+2)],
\eeq
which equals 1 if there are infinitely many twin primes, zero otherwise. 

It may turn out that a halting algorithm for computing $T$ exists, and that we humans simply have not discovered it yet.
In contrast, statements in the formal system that are undecidable statements in G\"odel's first incompleteness theorem 
correspond to, in this mathematical structure, 
predicates similar to that of \eq{TwinPrimeEq2} that cannot be computed by {\it any} halting algorithm, and are thus undefined in a computational sense.
Despite this, these Boolean functions are similar to \eq{TwinPrimeEq2} in being mere statements about integers;
the twist is that the quantifiers are used in an
insidious way that can be interpreted as self-referencing, so that any halting computation of the function would lead to a contradiction.

This is intimately linked to the famous halting problem in computer science \cite{Church,Turing}.
The proofs by Church and Turing that no halting algorithm can determine whether an arbitrary program will halt uses an analogous
self-reference and quantification over all programs to reach its conclusion.
Indeed, a slightly weaker form of G\"odel's first incompleteness theorem is an easy consequence of the 
undecidability 
of the halting problem. 

The results of G\"odel, Church and Turing thus show that under certain circumstances, 
there are questions that can be posed but not answered\footnote{
Note that what is striking about G\"odel's first incompleteness theorem is not the existence of undecidable statements {\it per se}.
For example, if one takes a simple decidable formal system such as Boolean algebra or Euclidean geometry and drops one of the axioms, 
there will be statements that can neither be proved nor refuted.
  % As a trivial example, in a formal system containing no axioms, any statement (WFF) is both unprovable and unfalsifiable.
  % Poor example since negation isn't defined, and  a bit lame to merely introduce it as a symbol.
A first remarkable aspect of the theorem is that even statements within the heart of standard mathematics (number theory) are undecidable. 
A second impressive fact is that one cannot merely blame weak axioms and inference rules for the failure, since 
the incompleteness theorem applies to {\it any} sufficiently powerful formal system.
In other words, adding more firepower does not help: 
if you add more tools to allow you to answer the undecidable questions, these same tools allow you to pose new questions that you cannot answer.
}.
We have seen that for a mathematical structure,
this corresponds to relations that are unsatisfactorily defined in the sense 
that they cannot be implemented by computations that are guaranteed to halt.

\subsection{Levels of mathematical relatity}
\label{MathRealityLevels}

So how well-defined do mathematical structures need to be to be real, \ie, members of the Level IV multiverse?
There is a range of interesting possibilities for what structures qualify:
\begin{enumerate}
\item No structures (\ie, the MUH is false).
\item Finite structures. These are trivially computable, since all their relations can be defined by finite look-up tables.
\item Computable structures (whose relations are defined by halting computations).
\item Structures with relations defined by computations that are not guaranteed to halt (\ie, may require infinitely many steps), 
like the example of \eq{TwinPrimeEq}.
Based on a G\"odel-undecidable statement, one can even define a function which is guaranteed to be uncomputable, 
yet would be computable if infinitely many computational steps were allowed. 
\item  Still more general structures.
       For example, mathematical structures with uncountably many set elements (like the continuous space examples in \Sec{SumUnitDimlessSec} and 
       virtually all current models of physics) are all uncomputable: one cannot even input the function arguments into 
       the computation, since even a single generic real number requires infinitely many bits to describe.
\end{enumerate}
The CUH postulates that (3) is the limit. If the CUH is false, then an even more conservative hypothesis is the 
Computable Finite Universe Hypothesis (CFUH) that (2) is the limit.

It is interesting to note that closely related issues have been hotly debated among mathematicians without any reference to physics
(see \cite{Feferman,Hersh} for recent updates). 
According to the finitist school of mathematicians (which included Kronecker, Weyl and Goodstein \cite{Goodstein}), 
representing an extreme form of so-called intuitionism and constructivism, 
a mathematical object does not exist unless it can be constructed from natural numbers in a finite number of steps. This leads directly to (3).

\subsection{Conceptual implications of the CUH}

According to the CUH, the mathematical structure that is our universe is computable and hence well-defined in the strong sense that all 
its relations can be computed. There are thus no physical aspects of our universe that are uncomputable/undecidable, eliminating the 
above-mentioned concern that G\"odel's work makes it somehow incomplete or inconsistent.
In contrast, it remains unclear to what extent the above-mentioned alternatives (4) and (5) can be defined in a rigorous and self-consistent way.

By drastically limiting the number of mathematical structures to be considered, the CUH also removes potential paradoxes related to the Level IV multiverse.
A computable mathematical structure can by definition be specified by a finite number of bits. For example, each of the finitely many generating relations
can be specified as a finite number of characters in some programming language supporting arbitrarily large integers 
(along the lines of the Mathematica, Maple and Matlab packages). Since each finite bit string can be interpreted as an integer in binary, there are 
thus only countably many computable mathematical structures. 
The full Level IV multiverse (the union of all these countably infinitely many computable mathematical structures) is 
then {\it not} itself a computable mathematical structure, since it has 
infinitely many generating relations. The Level IV multiverse is therefore not a member of itself, precluding Russell-style paradoxes, which addresses a concern 
raised in \cite{McCabe06,PageNonsense}.

\subsection{Physics implications of the CUH}
\label{CUHmeasureSec}

Many authors have puzzled over why our physical laws appear relatively simple.
For example, we could imagine replacing the standard model gauge group 
$SU(3)\times SU(2)\times U(1)$ by a much more complicated Lie group, and more generally reducing 
the symmetries of nature such that way more than 32 dimensionless parameters \cite{axions} were required to describe the laws.
It is tempting to speculate that the CUH contributes to this relative simplicity by 
sharply limiting the complexity of the Level IV multiverse.
By banishing the continuum altogether, perhaps the CUH may also help downsize the inflationary landscape 
\cite{Bousso00,Feng00,KKLT03,Susskind03,AshikDouglas04},
and resolve the cosmological measure problem, which is in large part linked to the ability of a true continuum 
to undergo exponential stretching forever, producing infinite numbers of observers \cite{inflation,Easther06,Bousso06,Vilenkin06,Aguirre06}.
It is unclear whether some sort of measure over the Level IV multiverse is required to fully resolve the measure problem, but if this is the case 
and the CUH is correct, then the measure could depend on the algorithmic complexity of the mathematical structures, which would be finite.
Labeling them all by finite bit strings $s$ interpreted as real numbers on the unit interval $[0,1)$ 
(with the bits giving the binary decimals), the most obvious measure for a given structure $S$ would be the fraction of the unit interval covered by real numbers
whose bit strings begin with strings $s$ defining $S$.
A string of length $n$ bits thus gets weight $2^{-n}$, which means that the measure rewards simpler structures. The analogous measure for computer programs is advocated
in \cite{Schmidhuber00}. A major concern about such measures is of course that they depend on the choice of representation of structures or computations as
bit strings, and no obvious candidate currently exists for which representation to use.

As long as the number of bits required to describe our (bird's view) mathematical structure is smaller than required to describe our (frog's view) 
observed universe, the mathematical structure must contain parallel universes as discussed in \Sec{ComplexWorldSec}.
In particular, if the frog's view description involves any real numbers, parallel universes are guaranteed, 
since their description requires infinitely many bits while CUH guarantees that the mathematical structure can be described by finitely many bits.

Penrose has suggested that the laws of physics must involve mathematics triggering G\"odel's incompleteness theorems because we humans are 
able to prove theorems regarding undecidable statements {\etc}
However, the fact that we can talk about the uncomputable clearly does not violate the consistency of our universe being computable
(this point is also emphasized in \cite{Schmidhuber00}).
  
Barrow has suggested that if our universe is a computer program, then all 
the laws of physics must involve approximately computable functions \cite{BarrowTOE}. 
As emphasized by \cite{McCabe05}, this would only be required if the computer
is computing the time-evolution. It would not necessarily be required if the computer merely stores
the 4-dimensional history or defines it in some other way, say by computationally implementing all relations.

\subsection{Challenges for the CUH}

Above we focused on attractive features of the CUH such as 
ensuring that the Level IV multiverse is rigorously defined 
and perhaps mitigating the cosmological measure problem by limiting what exists.
However, the CUH also poses serious challenges that need to be resolved.

A first concern about the CUH is that it may sound like a surrender of the philosophical high ground, 
effectively conceding that although all possible mathematical structures are ``out there", some have privileged status.
However, my guess is that if the CUH turns out to be correct, if will instead be because the rest of the mathematical landscape 
was a mere illusion, fundamentally undefined and simply not existing in any meaningful sense.

Another challenge involves our definition of a mathematical structure as an equivalence class of descriptions.
Since the problem of establishing whether two halting computations define the same function may be uncomputable, 
we lack a general algorithm for establishing whether two descriptions belong to the same equivalence class.
If this matters in any observable way, it would presumably be via the measure problem, in case it makes a difference whether 
two equivalent descriptions are counted separately or only once.
The more conservative CFUH mentioned in \Sec{MathRealityLevels} is free from this ambiguity, since a simple halting 
algorithm can determine whether any two finite mathematical structures are equivalent.

A more immediate challenge is that virtually all historically successful theories of physics 
violate the CUH, and that it is far from obvious whether a viable computable alternative exists. 
The main source of CUH violation comes from incorporating the continuum, usually in the form of real or complex numbers, 
which cannot even comprise the input to a finite computation since they generically require infinitely many bits to specify.
Even approaches attempting to banish the classical spacetime continuum by discretizing or quantizing it tend to maintain 
continuous variables in other aspects of the theory such as field strengths or wave function amplitudes.
Let us briefly discuss two different approaches to this continuum challenge: 
\begin{enumerate}
\item Replacing real numbers by a countable and computable pseudo-continuum such as algebraic numbers.
\item Abandoning the continuum altogether, deriving the apparent physical continuum as an effective theory in some appropriate limit.
\end{enumerate}

In the first approach, algebraic numbers and algebraic functions\footnote{Here we use the phrase ``algebraic numbers'' as shorthand for the special case
of algebraic numbers over the field of rational numbers.}
are perhaps the best contenders.
There are only countably many of them, all defined by polynomials with integer coefficients.
An algebraic number $x$ is defined by a univariate polynomial $P$ through $P(x)=0$, 
a univariate algebraic function $y(x)$ is defined by a bivariate polynomial through $P(x,y)=0$, 
a bivariate algebraic function $z(x,y)$ is defined by a trivariate polynomial through $P(x,y,z)=0$, 
and so on.
Algebraic functions have some very attractive features.
They are closed under addition, multiplication, composition, differentiation, and even equation solving.
Moreover, all of these operations on algebraic functions, as well as Boolean-valued functions like $=$ and $<$, can be implemented by halting computations, making them a 
computable mathematical structure. 
In contrast, rational numbers and functions are not closed under equation solving ($x^2=2$ having no rational solution).
In the broader class of approximately computable functions, which includes algebraic functions, $\pi$, $\sin x$ and virtually all other constants and functions 
taught in introductory mathematics courses, the Boolean-valued functions $=$ and $<$ cannot be implemented by computations that are guaranteed to halt.
Algebraic functions are, however, {\it not} closed under integration, which wreaks havoc with attempts to solve the differential equations of classical physics.
It is also unclear how to formulate gauge invariance without the exponential function.

In the second approach, one abandons the continuum as fundamental and tries to recover it as an approximation.
We have never measured anything in physics to more than about 15 significant digits, and no experiment has been carried out whose outcome depends
on the hypothesis that a true continuum exists, or hinges on Nature computing something uncomputable \footnote{We can of 
course program a computer to check if there are infinitely many twin primes, say,
but if this is provably undecidable, then it would require an infinite amount of time (and memory) to provide an answer 
}
It is striking that many of the continuum models of classical mathematical physics (like the wave equation, 
diffusion equation and Navier-Stokes equation for various media) are known to be mere approximations of an underlying discrete collection of atoms.
Quantum gravity research suggests that even classical spacetime breaks down on very small scales. 
We therefore cannot be sure that quantities that we still treat as continuous (like the metric, field strengths, and quantum amplitudes) are not mere approximations of something discrete.
Indeed, difference equations, lattice models and other discrete computable structures can approximate our continuum physics models so well that we use them for numerical computations,
leaving open the questions of whether the mathematical structure of our universe is more like the former or more like the latter.
Some authors have gone as far as suggesting that the mathematical structure is both computable and finite like a cellular automaton \cite{Wolfram02,Schmidhuber00}.
Adding further twists, physics has also produced examples of how something continuous (like quantum fields) can produce a discrete solution (like a crystal lattice) which in
turn appears like a continuous medium on large scales, which in turns has effective discrete particles (say phonons). 
Such effective particles may even behave like ones in our standard model \cite{Wen05,LevinWen05}, raising the possibility that we may have multiple layers of effective 
continuous and discrete descriptions on top of what is ultimately a discrete computable structure.

\section{Conclusions}
\label{ConcSec}

This paper has explored the implications of the Mathematical Universe Hypothesis (MUH) 
that our external physical reality is a mathematical structure (a set of abstract entities with relations between them).
I have argued that the MUH follows from the external reality hypothesis (ERH) that there exists an external physical reality completely 
independently of us humans, and that it constitutes the opposite extreme of the Copenhagen interpretation and other  
``many words interpretations'' of physics where human-related notions like observation are fundamental.

\subsection{Main results}

In \Sec{ScratchSec}, we discussed the challenge of deriving our perceived everyday view (the ``frog's view'') of our world 
from the formal description (the ``bird's view'') of the mathematical structure,
and argued that although much work remains to be done here, promising first steps include
computing the automorphism group and its subgroups, orbits and irreducible actions. 
We discussed how the importance of physical symmetries and irreducible representations emerges naturally, 
since any symmetries in the mathematical structure correspond to physical symmetries,
and relations are potentially observable.
The laws of physics being invariant under a particular symmetry group 
(as per Einstein's two postulates of special relativity, say)
is therefore not an input but rather a
logical consequence of the MUH.
We found it important to define mathematical structures precisely, and concluded that 
only dimensionless quantities (not ones with units) can be real numbers.

In \Sec{ICsec}, we saw that since the MUH leaves no room for arbitrariness or fundamental randomness,
it raises the bar for what constitutes an acceptable theory, 
banishing the traditional notion of unspecified initial conditions.
This makes it challenging to avoid some form of multiverse. 
Our frog's view of the world appears to require a vast amount of information (perhaps $10^{100}$ bits) to describe.
If this is correct, then any complete theory of everything expressible with fewer bits must describe a multiverse.
As discussed in \Sec{PUsec}, to describe the Level I multiverse of Hubble volumes emerging from inflation with all 
semiclassical initial conditions may require only of order $10^3$ bits (to specify the 32 dimensionless parameters tabulated
in \cite{axions} as well as some mathematical details like the $SU(3)\times SU(2)\times U(1)$ symmetry group --- the rest of
those Googol bits merely specify which particular Hubble volume we reside in.
To describe the larger Level II multiverse may require even less information (say $10^2$ bits
to specify that our mathematical structure is string theory rather than something else).
Finally, the ultimate ensemble of the Level IV multiverse would require $0$ bits to specify, since it has no free parameters.
Thus the algorithmic complexity (information content) of a multiverse is not only smaller than for the sum of its parts, but even 
smaller than for a generic {\it one} of its parts.
Thus in the context of the Level I multiverse, eternal inflation 
explains not only why the observed entropy of our universe is so low, but also why it is so high.

Staying on the multiverse topic, we defined and extensively discussed the Level IV multiverse of mathematical structures.
We argued that the existence of this Level IV multiverse (and therefore also Levels I, II and III as long as inflation and quantum mechanics
are mathematically consistent) is a direct consequence of the MUH.

In \Sec{SimSec}, we revisited the widely discussed idea that our universe is some sort of computer simulation.
We argued that, in the MUH context, it is unjustified to identify the 1-dimensional computational sequence with our 1-dimensional time,
because the computation needs to {\it describe} rather than {\it evolve} the universe.
There is therefore no need for such computations to be run.

In \Sec{GoedelSec}, we explored  how mathematical structures, formal systems, and computations are closely related, 
suggesting that they are all aspects of the same
transcendent structure (the Level IV multiverse) whose nature we have still not fully understood.
We explored an additional assumption: the Computable Universe Hypothesis (CUH)
that the mathematical structure that is our external physical reality is
defined by computable functions. We argued that this assumption may be needed tor the MUH to make sense, 
as G{\"o}del incompleteness and Church-Turing uncomputability will otherwise correspond to 
unsatisfactorily defined relations in the mathematical structure that require infinitely many computational steps to evaluate.
We discussed how, although the CUH brings severe challenges for future exploration, it may help with the cosmological measure problem

\subsection{The MUH and the Philosophy of science}

If the MUH is correct, then a number of hotly debated issues in the philosophy of science are cast in a different light.

As mentioned in \Sec{ScratchSec}, the conventional approach holds that a 
theory of mathematical physics can be broken down into (i) a mathematical structure, (ii) an empirical domain 
and (iii) a set of correspondence rules which link parts of the mathematical structure with parts of the empirical 
domain. If the MUH is correct, then (ii) and (iii) are redundant in the sense that they can, at least in principle, be derived from (i). 
Instead, they can be viewed as a handy user's manual for the theory defined by (i).

Since the MUH constitutes an extreme form of what is known as structural realism in the philosophy literature 
\cite{Ladyman98,McCabe06}, let us briefly comment on how it addresses three traditional criticisms of structural realism
\cite{Pooley05,McCabe06}.

The first issue is known as ``Jones under-determination", and occurs when  
different formulations of a theory involve different mathematical structures \cite{Jones91,Pooley05}, 
yet are empirically equivalent.
In the above-mentioned philosophical context with three theoretical domains, it has been questioned to what extent the mathematical 
structure of our world can be said to be real if it might not is be unique.
From the MUH perspective, such non-uniqueness is clearly not a problem, as the mathematical structure alone defines the theory.
If many different mathematical structures look the same to their inhabitants, this merely complicates the above-mentioned measure problem.

The second issue involves the possibility of a theory with a fixed mathematical structure possessing more than one realist 
interpretation \cite{Pooley05}. Since there is nothing ambiguous about domain (ii) and how observers perceive their world,
this is more an issue with traditional definitions of ``realist interpretation". 

The third issue, known as the ``Newman problem'', is of a more technical nature. As discussed in \cite{Pooley05},
the concern is that any structure can be embedded into others involving larger (typically infinite) cardinality, and
that this arguably collapses structural realism into empiricism, invalidating claims that science is in some way latching onto 
reality beyond the phenomena. With the definition of mathematial structures used in the present paper (using the left circle in \fig{StructureFig},
whereas Ladyman and others have emphasized the top circle), this becomes a non-issue. Replacing a mathematical structure by a larger 
one that contains it as a substructure may or may not have effects observable in the frog perspective (directly or via measure issues).
Either way, the mathematical structure --- domain (i) --- constitutes the theory. 

In summary, all of these issues hinge on differing conventions for what to mean by baggage words like ``theory'', ``realism'', etc.

\subsection{Outlook}

If the MUH is correct, if offers a fresh perspective on many hotly debated issues at the foundations of physics, mathematics and computer science.
It motivates further interdisciplinary research on the relations illustrated in \fig{StructureFig}.
This includes searching for a computable mathematical structure that can adequately approximate our current standard model of physics, dispensing with the continuum.

Is the MUH correct, making such efforts worthwhile?
One of its key predictions is that physics research will uncover mathematical regularities in nature.
As mentioned above, the score card has been amazing in this regard, ever since the basic 
idea of a mathematical universe was first articulated by the Pythagoreans, 
prompting awestruck endorsements 
of the idea 
from Galileo, Dirac \cite{Dirac31}, Wigner \cite{Wigner67} and others even before the 
standard models of particle physics and cosmology emerged.
I know of no other compelling explanation for this trend
other than that the physical world really is 
mathematical.

A second testable prediction of the MUH is that the Level IV multiverse exists, so that
out of all universes containing observers like us, we should expect to find ourselves in a rather typical one.
Rigorously carrying out this test requires solving the measure problem, \ie, computing 
conditional probabilities for observable quantities given other observations (such as our existence)
and an assumed theory (such as the MUH, or the hypothesis that only some specific mathematical structure like string theory 
or the 
Lie superalgebra $mb(3|8)$ \cite{Larsson01}
exists).
Further work on all aspects of the measure problem is urgently needed regardless of whether the MUH is correct, as this is necessary for
observationally testing any theory that involves parallel universes at any level, including cosmological inflation and 
the string theory landscape \cite{inflation,Easther06,Bousso06,Vilenkin06,Aguirre06}.
Although we are still far from understanding selection effects linked to the requirements for life, we can start
testing multiverse predictions by assessing how typical our universe is as regards dark matter, dark
energy and neutrinos, because these substances affect only better understood processes like galaxy
formation. 
Early such tests have suggested (albeit using questionable assumptions) 
that the observed abundance of these three substances is indeed rather typical of what you might 
measure from a random stable solar system in a multiverse where these abundances vary from universe to universe
\cite{BarrowTipler,LindeLambda,Weinberg87,Linde88,anthroneutrino,anthrolambdanu,axions}.

It is arguably worthwhile to study implications of the MUH even if one subscribes to an alternative viewpoint, as it forms a logical extreme in a broad spectrum 
of philosophical interpretations of physics.
It is arguably extreme in the sense of being maximally offensive to human vanity.
Since our earliest ancestors admired the stars, our human egos have suffered a series of blows. For starters, we are smaller than we thought. 
Eratosthenes showed that Earth was larger than millions of humans, and his Hellenic compatriots realized that the solar system was thousands of times larger still. 
Yet for all its grandeur, our Sun turned out to be merely one rather ordinary star among hundreds of billions in a galaxy that in turn is 
merely one of billions in our observable universe, the spherical
region from which light has had time to reach us during the 14 billion years since our big bang. Then there are more (perhaps infinitely many) such regions. 
Our lives are small temporally as well as spatially: if this 14 billion year cosmic history were scaled to one year, then 100,000 years of human history 
would be 4 minutes and a 100 year life would be 0.2 seconds. 
Further deflating our hubris, we have learned that we are not that special either. Darwin taught us that we are animals, Freud taught us
that we are irrational, machines now outpower us, and just last year, Deep Fritz outsmarted our Chess champion Vladimir Kramnik. 
Adding insult to injury, cosmologists have found that we are not even made out of the majority substance.
The MUH brings this human demotion to its logical extreme: not only is the Level IV Multiverse larger still, but even the languages, the notions
and the common cultural heritage that we have evolved is dismissed as ``baggage'', stripped of any fundamental status for describing the ultimate reality.

The most compelling argument {\it against} the MUH
hinges on such emotional issues: it arguably feels counterintuitive and disturbing.
On the other hand, placing humility over vanity has proven a more fruitful approach to physics, as emphasized by Copernicus, Galileo and Darwin.
Moreover, if the MUH is true, then it constitutes great news for science, 
allowing the possibility that an
elegant unification of physics, mathematics and computer science
will one day allow us humans to understand our reality even more deeply than many dreamed would be possible.

\bigskip
{\bf Acknowledgements:}
The author wishes to thank Anthony Aguirre, Ang{\'e}lica de Oliveira-Costa, Greg Huang, Kirsten Hubbard, Bihui Li, Gordon McCabe, 
Bruno Marchal, George Musser, 
David Raub, Ross Rosenwald, J\"urgen Schmidhuber, Jan Schwindt, Lee Smolin, John Tromp, David Vogan, Charles Waldman and especially 
Harold S.~Shapiro for helpful comments, and 
the John Templeton Foundation for sponsoring the symposium ``Multiverse and String Theory: Toward Ultimate Explanations in Cosmology''
held on 19-21 March 2005 at Stanford University, which inspired part of this work.
This work was supported by NASA grant NAG5-11099,
NSF grants AST-0134999 and AST-0708534,
a grant from the John Templeton foundation
and fellowships from the David and Lucile
Packard Foundation and the Research Corporation.   

\appendix

\section{What is a mathematical structure?}

This Appendix gives a discussion of mathematical structures at a more technical level, 
written primarily with physicists and mathematicians in mind.
The mathematics described here falls within the subject known as model theory. 
For an introductory textbook on the subject that is reasonably accessible to physicists, see \cite{Hodges}.

There are two common ways of introducing mathematical structures: as relations between abstract entities 
(as in \Sec{HypothesisSec}) or via formal systems (as in, \eg, \cite{toe}).
The two are intimately linked: in model theory terminology, the former is said to be a model of the latter.
Although I emphasized the latter in \cite{toe}, I have centered 
the discussion in this paper around the former, both for pedagogical reasons and 
because it is more closely linked to our discussion of issues such as symmetry and computability.

This Appendix is organized as follows. After a definition of the concept of a mathematical structure, 
this is illustrated with examples of a few simple finite ones.  
We then give examples of infinite ones, and discuss various subtleties related to units, observability, 
G\"odel-incompleteness and Church-Turing incomputability.

\subsection{Mathematical structures}
\label{StructureDefSec}

We define a  mathematical structure as a set $S$ of abstract entities and relations $R_1$, $R_2$, ... between 
them\footnote{Throughout this paper, I have tried to conform to the standard mathematical terminology as 
closely as possible without obscuring physical intuition. Unless explicitly specified, the usage is standard.
The mathematical structure definition given here differs from the common one of {\eg} \cite{Hodges} in two ways:
\begin{enumerate}
\item It allows multiple types (corresponding to the different sets $S_i$), much like different data 
types (Boolean, integer, real, \etc) in a programming language, because this greatly simplifies the definition of more 
complex structures cropping up in physics.
\item It defines a mathematical structure by {\it all} its relations. This means that one only needs to 
specify generators, a finite set of relations that generate all others.
\end{enumerate}
}.
Specifically, let us define the entities and relations of a mathematical structure as follows:\\
Given a finite number of sets $S_1$, $S_2$, ..., $S_n$, 
\begin{enumerate}
\item the set of entities is the union $S=S_1\union S_2\union...\union S_n$,
\item the relations are functions on these sets, specifically mappings from some number of sets to a set:
$S_{i_1}\times S_{i_2}\times ...\times  S_{i_k}\mapsto S_j$.
\end{enumerate}
For a Boolean-valued function $R$, (a function whose range $S_j=\{0,1\}$) we
say that ``the relation $R$ holds'' 
for those variables where $R=1$.
Functions that are not Boolean-valued can readily be interpreted as relations as well --- for instance,
we say that the relation $R(a,b)=c$ holds if the function $R$ maps $(a,b)$ to $c$.\footnote{The 
reader preferring to start without 
assuming the definition of a function can begin with Zermelo-Fraenkel set theory alone
and define relations as subsets of product sets of the type $S_{i_1}\times S_{i_2}\times ...\times  S_{i_k}$.
Subsets are equivalent to Boolean functions:
one says that the relation $R(a_1,a_2,...,a_n)$ holds 
(and that the Boolean function $R(a_1,a_2,...,a_n)$ is true) if $(a_1,a_2,...,a_n)\in R$, false otherwise.
Non-Boolean functions are definable by
interpreting a relation $R$ as an $S_{i_k}$-valued function and writing  
$R(a_1,a_2,...,a_{n-1})=a_n$
if for any values of the first $n-1$ variables, there is one and only one value $a_n$ of the last variable such that 
the relation $R(a_1,a_2,...,a_n)$ is true.
}
A mathematical structure typically has a countably infinite number of relations.
It is defined by
\begin{enumerate}
\item specifying the entities,
\item specifying a finite set of relations (``generating relations''), and
\item the rule that composition of any two relations generates a new relation. 
\end{enumerate}
This means that there are typically infinitely many equivalent ways of defining a given mathematical structure,
the most convenient one often being the shortest. 
We say that
one mathematical structure definition generates another if its generating relations generate all generating relations of the other, so two 
structure definitions are equivalent if they generate each other. 
There is a simple halting algorithm for determining whether any two finite mathematical structure definitions are equivalent.

\subsection{Mathematical structures examples}

Let us clarify all of this with a few examples, starting with {\it finite} mathematical structures (ones where there are only
a finite number of sets $S_i$ with a finite number of elements).

\subsubsection{Example: Boolean algebra}
\label{BooleanExampleSec}

This example corresponds to the case of a single set with only two elements, with the generating relations
indicated by the following tableaus: 
\beqa{BooleanStructureEq1}
S&=&
\begin{tabular}{|c|}
\hline
$0$\\
$1$\\
\hline
\end{tabular}\,,
\quad
R_1=
\begin{tabular}{|c|}
\hline
$0$\\
\hline
\end{tabular}\,,
\quad
R_2=
\begin{tabular}{|c|}
\hline
$1$\\
\hline
\end{tabular}\,,
\quad
R_3=
\begin{tabular}{|c|}
\hline
$1$\\
$0$\\
\hline
\end{tabular}\,,
\quad
R_4=
\begin{tabular}{|cc|}
\hline
$0$&$0$\\
$0$&$1$\\
\hline
\end{tabular}\,,
\nonumber\\
R_5&=&
\begin{tabular}{|cc|}
\hline
$0$&$1$\\
$1$&$1$\\
\hline
\end{tabular}\,,
\quad
R_6=
\begin{tabular}{|cc|}
\hline
$1$&$1$\\
$0$&$1$\\
\hline
\end{tabular}\,,
\quad
R_7=
\begin{tabular}{|cc|}
\hline
$1$&$0$\\
$0$&$1$\\
\hline
\end{tabular}\,,
\quad
R_8=
\begin{tabular}{|cc|}
\hline
$1$&$1$\\
$1$&$0$\\
\hline
\end{tabular}\,.
\eeqa
The $2\times 2$ boxes specify the values of functions with two arguments analogously to the multiplication 
table of \eq{C2StructureEq}, and the smaller boxes analogously define functions taking zero or one arguments.
These eight relations are normally denoted by the symbols 
$F$, $T$, $\sim$, $\&$, $\bigvee$, $\Rightarrow$, $\Leftrightarrow$ and $|$, respectively, and referred
to as ``false'', ``true'', ``not'', ``and'', ``or'', ``implies'', ``is equivalent to (xor)'' and ``nand''.

As an illustration of the idea of equivalent definitions, note that 
\beq{BooleanStructureEq2}
S=
\begin{tabular}{|c|}
\hline
$0$\\
$1$\\
\hline
\end{tabular},
\quad
R =
\begin{tabular}{|cc|}
\hline
$1$&$1$\\
$1$&$0$\\
\hline
\end{tabular}
\quad
\eeq
defines the {\it same} mathematical structure as \eq{BooleanStructureEq1}. This is because 
the single relation $R$ (``nand'', also known as Sheffer symbol $|$) generates all the relations of \eq{BooleanStructureEq2}:
\beqa{ShefferEquivalenceEq1}
R_1&=&R(R(X,R(X,X)),R(X,R(X,X))),\nonumber\\ 
R_2&=&R(X,R(X,X)),\nonumber\\ 
R_3(X)&=&\nonumber R(X,X),\\ 
R_4(X,Y)&=&R(R(X,Y),R(X,Y)),\nonumber\\ 
R_5(X,Y)&=&R(R(X,X),R(Y,Y)),\nonumber\\ 
R_6(X,Y)&=&R(X,R(Y,Y)),\nonumber\\ 
R_7(X,Y)&=&R(R(X,Y),R(R(X,X),R(Y,Y))),\nonumber\\ 
R_8(X,Y)&=&R(X,Y),\nonumber
\eeqa
or, in more familiar notation,
\beqa{ShefferEquivalenceEq2}
F			&=&(X|(X|X))|(X|(X|X)),\nonumber\\ 
T			&=&X|(X|X),\nonumber\\ 
\sim X			&=&\nonumber X|X,\\ 
X\&Y			&=&(X|Y)|(X|Y),\nonumber\\ 
X\bigvee Y		&=&(X|X)|(Y|Y),\nonumber\\ 
X\Rightarrow Y		&=&X|(Y|Y),\nonumber\\ 
X\Leftrightarrow Y	&=&(X|Y)|((X|X)|(Y|Y)),\nonumber\\ 
X|Y			&=&X|Y.\nonumber
\eeqa
(Note that the parentheses are needed since, whereas $\&$, $\bigvee$ and $\Leftrightarrow$ 
are associative, $|$ is not.)
Not or (``nor'') generates the same mathematical structure as ``nand''.
Another popular choice of generators for Boolean algebra, employed by G\"odel, is using only
$R_3$ and $R_5$ ($\sim$ and $\bigvee$) from \eq{BooleanStructureEq1}.

Not only the relations of \eq{BooleanStructureEq1},
but {\it all} possible Boolean functions of arbitrarily many (not just zero, one or two) 
variables can be expressed in terms of the Sheffer relation $|$ alone.
This means that the mathematical structure of Boolean algebra 
is simply all Boolean functions.

This simple example also illustrates why we humans create ``baggage'', symbols and words with implied meaning, 
like ``or''.
Of all of the infinitely many equivalent definitions of Boolean algebra, 
\Eq{BooleanStructureEq2} is clearly the most economical, with only a single generating relation.
Why then is it so common to define symbols and names for more? Because this provides convenient 
shorthand and intuition when working with Boolean algebra in practice.
Note that we humans have come up with familiar names 
for {\it all} nonary, unary and binary relations:
There are $2^{2^0}=2$ nonary ones ($F$ \& $T$), 
$2^{2^1}=4$ unary ones 
($F$, $T$, the identity relation and $\sim$, the last one being the only one
deserving a new symbol)
and $2^{2^2}=16$ binary ones ($F$, $T$, $X$, $\sim X$, $Y$, $\sim Y$,
$X\& Y$, 
$\sim(X\& Y)$, $X\bigvee Y$, $\sim(X\bigvee Y)$, 
$X\Leftrightarrow Y$, $\sim(X\Leftrightarrow Y)$,
$X\Rightarrow Y$, $\sim(X\Rightarrow Y)$,
$Y\Rightarrow X$, $\sim(Y\Rightarrow X)$.

\subsubsection{Example: the group $C_3$}

As another very simple example of a finite mathematical structure, consider the 3-element group defined by 
\beq{Z3StructureEq}
S=
\begin{tabular}{|c|}
\hline
$0$\\
$1$\\
$2$\\
\hline
\end{tabular},
\quad
R_1=
\begin{tabular}{|c|}
\hline
$0$\\
\hline
\end{tabular},
\quad
R_2=
\begin{tabular}{|c|}
\hline
$0$\\
$2$\\
$1$\\
\hline
\end{tabular},
\quad
R_3=
\begin{tabular}{|ccc|}
\hline
$0$&$1$&$2$\\
$1$&$2$&$0$\\
$2$&$0$&$1$\\
\hline
\end{tabular}.
\eeq
The usual notation for these relations is $R_1=e$, $R_2(g)=g^{-1}$, $R_3(g_1,g_2)=g_1 g_2$.
In the above-mentioned Boolean algebra example, there were many equivalent choices of generators.
The situation is similar for groups, which can be defined with a single generating relation
$R(a,b) = a b^{-1}$, in terms of which the three relations above can all be reexpressed:
\beqa{GroupStructureEq2}
R_1		&=&R(a,a) = e,\nonumber\\ 
R_2(a)	&=&R(R(a,a),a) = a^{-1},\nonumber\\ 
R_3(a,b)	&=&R(a,R(R(b,b),b)) = a b.
\eeqa
The specific example of \eq{Z3StructureEq} thus has the following simpler equivalent definition:
\beq{Z3StructureEq3}
S=
\begin{tabular}{|c|}
\hline
$0$\\
$1$\\
$2$\\
\hline
\end{tabular},
\quad
R=
\begin{tabular}{|ccc|}
\hline
$0$&$2$&$1$\\
$1$&$0$&$2$\\
$2$&$1$&$0$\\
\hline
\end{tabular}
\eeq
Note that the identity element can be read off as the diagonal element(s) of this ``division table''
and that the inverses can be read off from the first row.

\subsubsection{Encoding finite mathematical structures}

Since the generating relations of a finite mathematical structure can be specified by simply tabulating 
all their values as in equations\eqn{BooleanStructureEq1} and\eqn{BooleanStructureEq2}, it is straightforward to encode 
the structure definition as a finite string of integers. Here is a simple example of such an encoding scheme:
$$\brack{\# of sets}\brack{\# of relations}\brack{set def.~1}...\brack{rel.~def.~1}...$$
Here each set definition is simply an integer giving the number of elements in 
the set.
Each relation (function) definition has the following structure:
$$\brack{\# of args}\brack{arg type 1}...\brack{output type}~\brack{value array}$$
For each of the arguments as well as for the output returned by the relation, 
the type is simply one of the previously defined sets, numbered 0, 1, 2, ....
The value tensor simply lists all the values as an array (0-dimensional, 1-dimensional, 2-dimensional, {\etc}
depending on the number of arguments, looping over earlier arguments first in the same way that computer languages such as 
Fortran store arrays. Elements of the output set are numbered 0, 1, ....

Table~{\StructureTable} illustrates how this works for a number of simple examples.
The two equivalent definitions of Boolean algebra corresponding to 
equations\eqn{BooleanStructureEq1} and\eqn{BooleanStructureEq2} correspond to the strings 
\beqa{StringExampleEq}
s_1&=&\{ 1 8 2 
0 0 0 
0 0 1 
1 0 0 0 1 
2 0 0 0 0 0 0 1 
2 0 0 0 0 1 1 1\nonumber\\
&&\>\>2 0 0 0 1 0 1 1  
2 0 0 0 1 0 0 1
2 0 0 0 1 1 1 0 
\},\nonumber\\
s_2&=&\{ 1 1 2 2 0 0 0 1 1 1 0 \}\nonumber\\
\eeqa
respectively; the latter is clearly more economical.

Such an explicit encoding scheme allows mathematical structures to be defined in a way more amenable to computers.
We say that an encoding $s_1$ {\it generates} an encoding $s_2$ if 
$s_1$ contains all the sets of $s_2$ and all relations in $s_2$ are generated by relations in $s_1$.
We say that two encodings $s_1$ and $s_2$ are {\it equivalent} 
if $s_1$ generates $s_2$ and $s_2$ generates $s_1$. In our example above, $s_1$ and $s_2$ are equivalent, since they generate each other.

If one wishes to catalog and enumerate all finite mathematical structures, it is convenient to order 
them by some measure of their complexity. There are many ways of doing this. 
For example, one could define 
the complexity $H(s)$ (measured in bits) of an encoding
$s=\{k_1,k_2, ...,k_n\}$
of a mathematical structure as
\beq{ComplexityDefEq}
H(s)=\sum_{i=1}^n\log_2(2+k_i).
\eeq
Then the integers 0, 1 and 2 add complexities of 1 bit, $\log_2(3)\approx 1.6$ bits and 2 bits, respectively,
and the two above encodings of Boolean algebra have complexities of $H(s_1)=72$ bits and $H(s_2) = 16$ bits, respectively,
whereas the empty set in Table~{\StructureTable} has 3.6 bits.

This allows us to explicitly define a finite mathematical structure as an equivalence class of encodings.
We label it by the encoding that has the lowest complexity, using lexicographical ordering to break any ties,
and we define the complexity of the mathematical structure as the complexity of this simplest encoding.
In other words, this provides an explicit scheme for enumerating all finite mathematical structures.

\subsubsection{Infinite mathematical structures}

The more general case of infinite mathematical structures is discussed at length in \Sec{GoedelSec}.
The key point is that there are only countably many computable ones (which include all the countably many finite structures), 
and that to enumerate them all, the encoding scheme 
above must be replaced by one that defines the relations not by explicit tabulation, but by providing an algorithm or computer 
program that implements them.

\end{document}